\documentclass[aps,prl,twocolumn,showpacs,superscriptaddress,floatfix]{revtex4-2}  
\usepackage{ucs}
\usepackage{natbib}
\usepackage{graphicx}  
\usepackage{dcolumn}   
\usepackage{bbold}
\usepackage{bm}        
\usepackage{amssymb}   
\usepackage{amsmath}
\usepackage{color}
\usepackage{verbatim}
\usepackage{mathrsfs}
\usepackage{ulem}
\usepackage{mathtools}
\usepackage[english]{babel}

\newcommand{\bg}{\begin{pmatrix}}
\newcommand{\ed}{\end{pmatrix}}

\newcommand{\red}[1]{\textcolor{red}{#1}}

\renewcommand{\vec}[1]{{\boldsymbol #1}}

\usepackage{cancel}
\usepackage{comment}
\usepackage{hyperref}

\usepackage{xcolor}
\hypersetup{
	colorlinks,
    linkcolor={magenta},
	citecolor={green!50!black},
	urlcolor={blue!50!black}
}

\begin{document}

\title{Imaging phase winding in topological superconductors with a fork-tip Josephson STM}
\author{Vladislav Poliakov}\email{vlad\_p@mit.edu}
\affiliation{Department of Physics, Massachusetts Institute of Technology, Cambridge, MA 02139, USA}
\author{Archisman Panigrahi}\email{archi137@mit.edu}
\affiliation{Department of Physics, Massachusetts Institute of Technology, Cambridge, MA 02139, USA}

\begin{abstract}

The topological phase winding in a chiral superconductor remains challenging to access by conventional real-space probes. We propose a scheme to probe this winding with a fork-tip Josephson STM near an impurity, where two superconducting tips form an interferometer sensitive to phase differences between two positions. The phase winding remains accessible via the fork-tip critical current even if the relative phase between the fork-tip and sample fluctuates, establishing a route towards a local phase-sensitive STM probe of topological superconductivity. We numerically compute the interference pattern to demonstrate that the magnitude and phase of the order parameter can be reconstructed as a function of angle, providing a route towards resolving the longstanding open experimental problem of directly detecting phase winding in topological superconductors.
\end{abstract}

\date{\today}

\maketitle

\textit{\color{magenta}Introduction---}
Probing the topological phase winding of superconducting order parameters in chiral superconductors remains a major experimental challenge~\cite{Kallin_chiral_2016,Sato_topological_2017,Read_paired_2000}. Conventional spectroscopic probes such as angle-resolved photoemission, thermal transport and nuclear magnetic resonance characterize the gap function anisotropy and the quasiparticle spectrum, but cannot access the phase~\cite{Damascelli_angle-resolved_2003, Pustogow_constraints_2019, Matsuda_nodal_2006}. Sign-changes of the order parameter can be detected by Josephson interferometry, first established for $d$-wave superconductors~\cite{Wollman_experimental_1993, VanHarlingen_phase-sensitive_1995, Tsuei_pairing_2000}, while broken time-reversal symmetry can be revealed by muon spin relaxation and polar Kerr effect~\cite{Luke_time-reversal_1998, Xia_high_2006}. However, these techniques cannot directly reconstruct the angular dependence of the phase of the gap function $\Delta_{\vec k} = |\Delta_{\vec k}| e^{i\varphi_{\vec k}}$, which remains an open experimental problem.


Scanning tunneling microscopy (STM) provides a natural route towards local measurements in unconventional superconductors where boundaries, impurities and vortices produce spatial variations that reflect the symmetry and phase winding of the superconducting gap function~\cite{Hoffman_imaging_2002, Hanaguri_coherence_2009, Fischer_scanning_2007,Rodrigo_scanning_2008,Engstrom_detecting_2025, Balatsky_impurity-induced_2006, Salkola_theory_1996, Pan_imaging_2000, Hudson_interplay_2001}. A conventional STM with a metallic tip probes the local density of states of quasiparticles, while a superconducting tip can additionally exhibit coherent tunneling of a Cooper pair from the tip to a superconducting sample, giving rise to a Josephson current at zero bias~\cite{Smakov_josephson_2001, Kimura_josephson_2009}. Josephson STMs featuring superconducting tips have been successfully used for precise measurements of spatial features such as vortices, Cooper pair density waves, as well as inhomogeneities in condensate density and order parameter~\cite{Smakov_josephson_2001, Bergeal_mapping_2008, Rodrigo_scanning_2008, Moreno_robust_2026,Kimura_josephson_2009,Kimura_scanning_2008, Hamidian_detection_2016, Cho_a_2019, Randeria_scanning_2016, Graham_imaging_2017}. For atomically sharp tips, the Josephson coupling energy is typically smaller than the temperature, placing the junction in the phase diffusive regime~\cite{Naaman_fluctuation_2001, Jack_critical_2016, Ast_sensing_2016}. Consequently, conventional single-tip Josephson STM can measure spatial variations of the local order parameter's magnitude, while the phase information is washed out by thermal fluctuations.

\begin{figure}[t]
    \centering
    \includegraphics[width=1.0\linewidth]{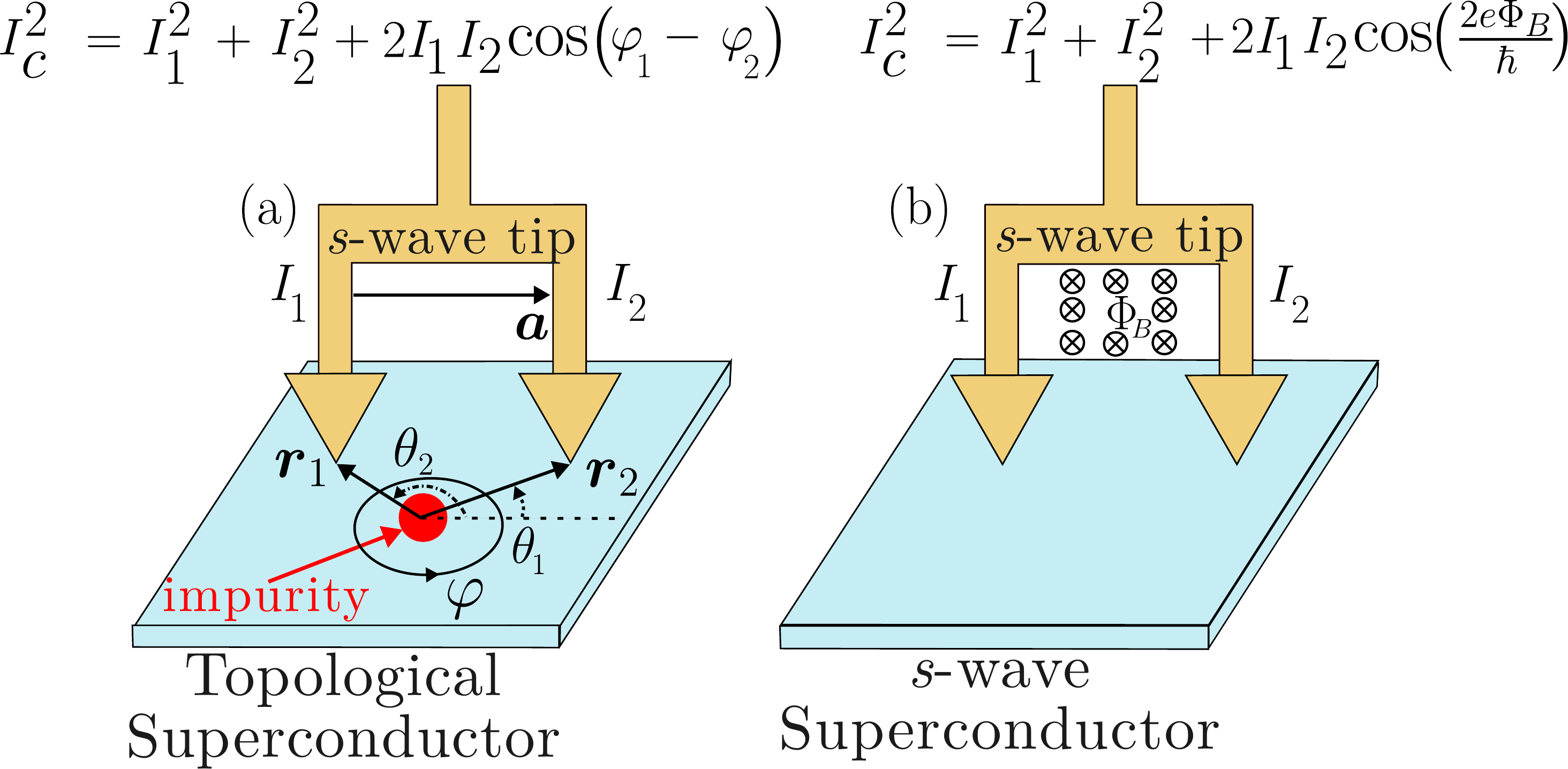}
    \caption{Schematic of (a) fork-tip Josephson STM near an impurity on a topological superconductor. The impurity generates a local phase winding in the anomalous Green's function around it, leading to the phase difference $(\varphi_1-\varphi_2)$ between two junctions and hence to the interference contribution to the critical current. (b) Conventional fork-tip interference in an $s$-wave superconductor where an Aharonov-Bohm phase difference is induced by an external magnetic flux $\Phi_B$.}
    \label{fig:fork-tip}
\end{figure}

To overcome this limitation, we propose replacing the single superconducting tip with a phase-coherent `fork-tip' consisting of two superconducting tips connected to the same electrode, forming a two-junction interferometer resembling a superconducting quantum interference device (SQUID) (see Fig.~\ref{fig:fork-tip}). Although the overall phase difference between the tip and the sample thermally fluctuates, since both the junctions share the same fluctuating phase, the interference retains the relative phase $\varphi_1 - \varphi_2$ of the anomalous Green's function at the two tip positions $\vec r_{1,2}$, resulting in a critical current $I_c$ given by
\begin{equation}\label{two-tip current general form}
    I_c^2(\vec{R})
= I_0^2(\vec{r}_1)+I_0^2(\vec{r}_2)
+2I_0(\vec{r}_1)I_0(\vec{r}_2)\cos(\varphi_1-\varphi_2).
\end{equation}
Here $I_0(\vec r_{1,2})$ are the critical Josephson currents of individual tips. In  conventional SQUID-like interferometry~\cite{Roychowdhury_a_2014, Roychowdhury_development_2014, Liao_simultaneously_2017, Liao_investigation_2019}, a similar phase difference can be generated by the Aharonov-Bohm effect associated with an external magnetic field (see Fig.~\ref{fig:fork-tip}(b)). In contrast, here the phase difference is intrinsically generated by the topological winding near an impurity.
Multi-tip STMs~\cite{Voigtlander_invited_2018, Nakayama_development_2012, Leeuwenhoek_modeling_2020} have been proposed and developed for measuring nonlocal transport~\cite{Szumniak_local_2026}, Aharonov-Bohm effect in an external magnetic field~\cite{Roychowdhury_a_2014, Roychowdhury_development_2014, Liao_simultaneously_2017, Liao_investigation_2019}, as well as for detecting spatial modulations~\cite{Coleman_triplet_2020} and majorana edge states~\cite{Ptok_probing_2020}. Nanoscale SQUIDs with proximity Josephson junctions have been recently realized in Ref.~\cite{Rog_tapping-mode_2026}, demonstrating the feasibility of integrating a Josephson junction device with STM tips. 

It was recently shown that the symmetry properties and the topological phase winding of an unconventional superconductor get imprinted in the tunneling conductance ($\frac{dI}{dV}$) near two impurities~\cite{Ding_hyperbolic_2023, Panigrahi_Poliakov_tomographic_2026}. The scattering of quasiparticles from two impurities or from an impurity and a reflective boundary~\cite{Panigrahi_Poliakov_particle_hole_2026} produces an interference signal in TDOS, which carries information about the symmetry of the superconducting order parameter. However, this contribution is parametrically weaker than the $\frac{dI}{dV}$ near a single impurity, which does not carry information about the phase winding, and therefore it must be isolated, for example, using Fourier filtering. In contrast, the fork-tip Josephson STM proposed here produces a phase-sensitive signal already in the first order in impurity strength, rather than at second order, and allows us to construct the angular dependence of the phase.

We propose a phase-coherent superconducting fork-tip STM that measures the Josephson current near an impurity on a superconductor, to probe the momentum dependent phase of its order parameter. We show that the real-space anomalous Green's function near a localized impurity acquires a correction that reflects the momentum-dependence of the order parameter. We derive the Josephson current near the impurity for a general momentum- and spin-dependent superconducting order parameter. The interference between the phase-coherent tips produces a characteristic spatial pattern carrying a unique fingerprint of the order parameter's angular dependence. We demonstrate that fitting these patterns allows us to reconstruct the angular dependence of both the gap-function magnitude $|\Delta_\theta|$ and phase $\varphi_{\theta}$ in $\vec k$-space.


\begin{figure*}
    \centering
    \includegraphics[width=0.89\linewidth]{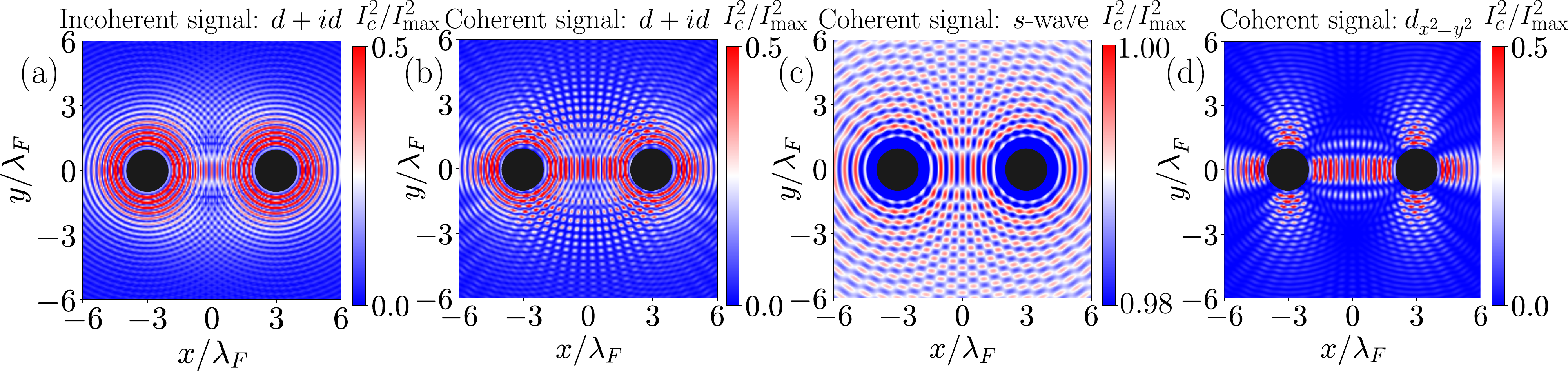}
    \caption{The spatial pattern of the squared fork-tip critical current near a single point impurity for different superconducting order parameters. (a) $I_c^2(\vec R)$ for $d+id$ superconductor ($\Delta_\theta=\tilde\Delta e^{2i\theta}$) without phase coherence between the two tips. Since there is no interference term ($I_c^2(\vec R)=I_{0}^2(\vec r_1) + I_{0}^2(\vec r_2)$) this signal is insensitive to the phase of the order parameter. Panel (b) shows the analogous squared critical current for two phase-coherent tips. The signal now exhibits hyperbolic fringes due to the interference term ($2I_0(\vec r_1)I_0(\vec r_2)\cos(\varphi_1-\varphi_2)$) in $I_c^2(\vec R)$. (c) Coherent signal for $s$-wave ($\Delta_\theta=\tilde\Delta$) superconductor, with a finite background due to the large bulk contribution to the anomalous Green's function $F(\vec r, \vec r, i\omega_n)$. (d) Coherent signal for $d_{x^2-y^2}$ ($\Delta_\theta=\tilde \Delta \cos(2\theta)$)  superconductor. Here, $I_c(\vec R)$ has a clear nodal structure along the nodal directions of the order parameter. For simplicity here we have set identical tunneling amplitude for both tips.
    }
    \label{fig:no-noise-plots}
\end{figure*}
\textit{\color{magenta} Superconducting STM current---}
We first derive the Josephson current in the weak tunneling limit for a general superconductor with arbitrary spin- and momentum-dependent order parameter. 
Unlike previous treatments concerning the Josephson current in an $s$-wave superconductor with spatially homogeneous anomalous Green's function~\cite{Graham_imaging_2017, Lauke_friedel_2018},  our general formulation retains arbitrary spin and momentum dependence of the gap function, as well as spatial inhomogeneity. This formulation is crucial for computing the impurity-induced correction to a superconductor with arbitrary spin and momentum dependent order parameter, as the Josephson current depends on the phase of the full anomalous Green’s function rather than on separate unperturbed and perturbed contributions.
Following Refs.~\cite{Larkin1967tunnel, Larkin1967boundary} we consider an arbitrary two-dimensional superconducting system coupled to an STM tip through a weak tunneling Hamiltonian,
\begin{equation}
\begin{aligned}\label{initial Hamiltonian}
\hat{H}&=\hat{H}_{\mathrm{tip}}+\hat{H}_{\mathrm{sys}}+\hat{H}_{\mathrm{t}} \\
\hat{H}_t&=\sum_{\alpha,\nu} \Big[t c^\dagger_{\alpha}(\vec r)d_\alpha(\nu) +t^* d^\dagger_{\alpha}(\nu) c_{\alpha}(\vec r) \Big].
\end{aligned}
\end{equation}
Here $c_\alpha(\vec r)$ annihilates an electron in the sample at position $\vec r$ with spin $\alpha$, and $d_\alpha(\nu)$ annihilates an electron in the tip mode $\nu$ with spin $\alpha$. 
The most general expression for the Josephson current for the superconductor in Eq.~\eqref{initial Hamiltonian} can then be written as (see Appendix A in Supplementary Materials (SM)~\cite{SM_fork-tip}),
\begin{equation}
\label{single tip Josephson through F}
I_J(\vec r) = -2e|t|^2 T \sum_{\mathclap{i\omega_n,\alpha\beta}} \operatorname{Im}\Big[ F^\dagger_{\alpha\beta}(\vec r,\vec r,i\omega_n) \sum_\nu F^{\mathrm{tip}}_{\alpha\beta}(\nu,i\omega_n) \Big],
\end{equation}
where the anomalous Green's function in the superconductor is defined as $F_{\alpha\beta}(\vec r, \vec r', i\omega_n) = -\int_0^\beta d\tau\, e^{i\omega_n\tau} \left\langle T_\tau \psi_\alpha(\vec r,\tau) \psi_\beta(\vec r',0) \right\rangle$~\cite{AGD}. 
If
the phase of the anomalous Green's function is independent of Matsubara frequency and spin indices,
\begin{eqnarray}\label{phase of F}
    F_{\alpha \beta}(\vec r, \vec r,i\omega_n)=|F_{\alpha \beta}(\vec r, \vec r,i\omega_n)| e^{i\varphi(\vec r)}.
\end{eqnarray}
then for a phase $\varphi_{\text{tip}}$ at the superconducting tip,
the Josephson current reduces to the conventional form, 
\begin{eqnarray}\label{single tip josephson}
    I_J(\vec r)=I_0(\vec r) \sin (\varphi(\vec r)-\varphi_{\rm tip}).
\end{eqnarray}

The relative phase $(\varphi-\varphi_{\rm tip})$ 
generates the Josephson energy $E=-E_J \cos (\varphi-\varphi_{\rm tip})$, whose strength is determined by the tunneling parameters. For an atomically sharp STM tip this energy is much smaller than the characteristic thermal energy ($E_J \ll k_BT_{\mathrm{eff}}$), placing the junction in the phase-diffusive regime~\cite{Ivanchenko1968, Ambegaokar_voltage_1969, Grabert_single_1992, Ingold_cooper-pair_1994},
where the relative phase fluctuates strongly, resulting in a vanishing average dc current at zero voltage bias. 
However, under a fixed current bias $I$, thermal fluctuations generate a finite average voltage $V$, and in the low-bias regime they are related by the Ivanchenko-Zil'berman formula~\cite{Ivanchenko1968},
\begin{eqnarray}\label{eq:I-V-relation-current-bias}
    I(V)=\frac{I_0^2 Z_{\mathrm{env}}}{2}\frac{V}{V^2+(\frac{2e}{\hbar}Z_{\mathrm{env}}k_BT_{\text{eff}})^2}.
\end{eqnarray}
Here $Z_{\mathrm{env}}$ and $T_{\mathrm{eff}}$ are parameters characterizing the thermal noise, specific to a particular device. By fitting the experimental $I$-$V$ curve, the Josephson current amplitude $I_0$ can be extracted. Consequently, this measurement involving a single tip probes the magnitude of the anomalous Green's function, but the phase $\varphi(\vec r)$ is completely lost due to thermal noise. Now we explain how the phase information is retained in fork-tip STM.

\textit{\color{magenta} Fork-tip STM---} We now extend the single tip analysis to the fork-tip STM where two tips are connected to the same superconducting electrode, forming a single V-shaped tip. Let the midpoint of the two tips be located at $\vec R$, and let $\vec a$ denote the vector connecting the two tips (see Fig.~\ref{fig:fork-tip}(a)), which probe the sample at the positions
\begin{eqnarray}
\vec r_1=\vec R+\frac{\vec a}{2},
\qquad
\vec r_2=\vec R-\frac{\vec a}{2}.
\end{eqnarray}

Since the tips are connected to the same phase-coherent superconducting electrode, they share the same phase $\varphi_{\mathrm{tip}}$. 
The total current through the two tips is,
\begin{eqnarray}\label{eq:I-J-R}
    I_J(\vec{R})=I_0(\vec{r}_1)\sin \left(\varphi_1-\varphi_{\mathrm{tip}}\right)+I_0(\vec{r}_2)\sin \left(\varphi_2-\varphi_{\mathrm{tip}}\right),
\end{eqnarray}
where $I_0(\vec{r}_{1,2})$ are the local critical Josephson currents 
at positions $\vec{r}_{1,2}$. 
$I_J(\vec{R})$ in Eq.~\eqref{eq:I-J-R} can be rewritten as,
\begin{eqnarray}
I_J(\vec{R})=I_c(\vec{R})\sin \left(\alpha(\vec{R})-\varphi_{\mathrm{tip}}\right)
\end{eqnarray}
where the critical current of the fork-tip junction $I_c(\vec R)$ is shown in Eq.~\eqref{two-tip current general form} and $\tan(\alpha) =\frac{ I_0(\vec{r}_1)\sin(\varphi_1)+I_0(\vec{r}_2)\sin(\varphi_2)}{ I_0(\vec{r}_1)\cos(\varphi_1)+I_0(\vec{r}_2)\cos(\varphi_2)
}$.
Once we introduce the thermal noise which affects the phase difference $(\alpha(\vec{R}) - \varphi_{\mathrm{tip}})$, the analysis is identical to a single-tip case, while $I_c^2(\vec R)$ defined in Eq.~\eqref{two-tip current general form} plays the role of critical current. Thus the experimentally accessible quantity is the critical current amplitude $I_c(\vec{R})$, which can be measured as a function of position.

According to Eq.~\eqref{single tip Josephson through F}, the local Josephson current amplitude can be obtained from the anomalous Green's function at positions $\vec{r}_{1,2}$,
\begin{equation}\label{Ic2 for fork tip general}
\begin{aligned}
I_c^2&(\vec R) \propto 
|\mathcal{F}(\vec r_1)+\mathcal{F}(\vec r_2)|^2  \\
=&\left|\mathcal{F}(\vec r_1)\right|^2
+
\left|\mathcal{F}(\vec r_2)\right|^2
+
2\left|\mathcal{F}(\vec r_1)\right|
\left|\mathcal{F}(\vec r_2)\right|
\cos\left(\varphi_1-\varphi_2\right),
\end{aligned}
\end{equation}
where
\begin{equation}\label{complex F}
\begin{aligned}
\mathcal{F}(\vec r_i)
&\equiv
    -2e |t|^2 T\sum_{i\omega_n,\alpha \beta}\;F_{\alpha \beta}^\dagger(\vec r_i, \vec r_i,i\omega_n) \sum_\nu F_{\alpha \beta}^{\mathrm{tip}}(\nu,i\omega_n), 
\end{aligned}
\end{equation}
which can be rewritten as $|\mathcal{F}(\vec r_i)|e^{i(\varphi_{\rm{tip}}-\varphi_i)}$.
Scanning the fork-tip STM across the sample, the interference of tunneling currents at positions $\vec{r}_{1,2}=\vec{R}\pm \vec{a}/2$ can be measured, which allows the real-space variation of phase $\varphi_i \equiv \varphi(\vec{r}_i)$ of the anomalous Green's function to be determined. Now we demonstrate that the momentum-dependence of the order-parameter's phase gets imprinted in the real-space phase of the anomalous Green's function near a localized potential impurity, enabling us to probe the angle dependence of the phase.

\textit{\color{magenta} Disorder induced anomalous Green's function---}
To demonstrate that the phase of the impurity-induced correction to anomalous Green's function mimics the momentum-dependent phase of the order parameter,
we start with a BdG Hamiltonian,
\begin{equation}\label{initial bdg}
\begin{aligned}
\hat H_0&=\frac{1}{2}\int \Psi_{\vec k}^\dagger H_0(\vec k) \Psi_{\vec k} \frac{d^2\vec k}{(2\pi)^2} \\
H_0(\vec k) &=
\begin{pmatrix}
\xi_{\vec k} & \Delta_{\vec k} \\
\Delta^*_{\vec k} & -\xi_{\vec k}
\end{pmatrix}
\end{aligned}
\end{equation}
with $\Psi_{\vec k} = (a_{\vec k} \quad a^\dagger_{-\vec k})^T$, where $a_{\vec k} = c_{\vec k, \uparrow}, a^\dagger_{-\vec k}=c^\dagger_{-\vec k,\downarrow}$ for a spin-singlet superconductor and $a_{\vec k} = c_{\vec k, \uparrow}, a^\dagger_{-\vec k}=c^\dagger_{-\vec k,\uparrow}$ for a spin-polarized superconductor.
An impurity placed at $\vec r_0$ generates a perturbation to the Hamiltonian,
\begin{eqnarray}
    \hat H_{\text{imp}}= \frac{1}{2}\int \Psi_{\vec r}^\dagger U\delta(\vec r-\vec r_0)\begin{pmatrix}
        1 & 0 \\
        0 & -1
    \end{pmatrix}
    \Psi_{\vec r} d^2 \vec r .
\end{eqnarray}

To study the impurity-induced corrections to the Josephson current, we compute the Green's function.
 The bare Matsubara Green's function is defined as $G_0(\vec k,i\omega_n)=(i\omega_n-H_0(\vec k))^{-1}$.
For a weak impurity strength $U$, the correction to the real space Green's function $G_0(\vec r, \vec r',i\omega_n) \equiv G_0(\vec r' - \vec r,i\omega_n)$ can be obtained with perturbation theory,
\begin{equation}\label{correction to G general formula}
\begin{aligned}
G(\vec r, \vec r',i\omega_n)
=G_0(\vec r'- \vec r,i\omega_n)+\delta G(\vec r,\vec r',i\omega_n)\\
\delta G(\vec r, \vec r',i\omega_n)=G_0(\vec r',i\omega_n)U\tau_z G_0(-\vec r,i\omega_n).
\end{aligned}
\end{equation}


The Josephson current is determined by the equal-point anomalous Green's function 
$F(\vec r, \vec r,i\omega_n) = \text{Tr}_{\tau}\left[\tau^{-}G(\vec r, \vec r,i\omega_n)\right]$ corresponding to the top-right element of the matrix $G$. The zeroth-order contribution 
to $F(\vec r, \vec r,i\omega_n)$ is obtained by taking a Fourier transform of $F_0(\vec k,i\omega_n)$ at $\vec r'=\vec r$,
\begin{eqnarray}\label{F0 for clean system}
    F_0(\vec r, \vec r,i\omega_n)=-\int\frac{\Delta_{\boldsymbol{k}}}{\omega_n^2+\xi_{\boldsymbol{k}}^2+|\Delta_{\boldsymbol{k}}|^2}\frac{d^2\vec k}{(2\pi)^2}.
\end{eqnarray}

A junction between a spin-singlet and a spin-triplet superconductor cannot facilitate any Josephson current, as the Cooper pairs in one of these systems cannot coherently tunnel to the other
(see Appendix A in (SM)~\cite{SM_fork-tip}).
For the rest of the paper, we focus on Josephson tunneling between a spin-singlet tip and a spin-singlet topological superconductor. Probing spin-triplet superconductors will require an STM with a Majorana mode~\cite{Lauke_friedel_2018, Kashuba_majorana_2017}, featuring a non-zero equal-point anomalous Green's function.
For an isotropic dispersion and at distances much larger than the Fermi wavelength ($r \gg \lambda_F$), the impurity-modified anomalous Green's functions can be computed semiclassically (see Appendix B in SM~\cite{SM_fork-tip}),
\begin{align}\label{eq:anomalous-green-phase-winding}
\delta F(\vec r,\vec r,i\omega_n)
&=
\frac{U k_F}{2\pi {{v_F^*}}^2}
\frac{\cos(2k_F r)}{r}
\frac{|\Delta_\theta| e^{i\varphi_{\theta}} e^{-\frac{2r\sqrt{\omega_n^2+|\Delta_\theta|^2}}{{v_F^*}}}}
{\sqrt{\omega_n^2+|\Delta_\theta|^2}}.
\end{align}
Here $k_F$ and ${v_F^*}$ are Fermi momentum and effective Fermi velocity. The impurity produces a correction $\delta F$ 
proportional to $|\Delta_{\theta}| e^{i\varphi_\theta}$, i.e.\ the order parameter at the momentum direction connecting the STM tip to the impurity. Thus, the impurity-induced anomalous Green's function in real space inherits the complex phase winding of $\Delta_{\vec k}$.

In chiral superconductors with approximately uniform gap magnitude, the equal-point anomalous Green's function vanishes in the bulk, since $\int \Delta_{\boldsymbol{k}} d\theta=0$.
The leading contribution to the Josephson current 
will therefore be given by the impurity-induced correction $\delta F$. Substituting the impurity-induced correction to the anomalous Green's function in Eqs.~\eqref{Ic2 for fork tip general} and~\eqref{complex F} allows calculating the fork-tip Josephson current. For a chiral superconductor the spatial pattern of the fork-tip critical current near the impurity will directly depend on the phase winding of the order parameter, whereas a single tip's critical current is independent of the topological phase winding.

\textit{\color{magenta}Reconstruction of the superconducting order parameter's angular dependence---}
Having established the general principle, we consider a simple case where the temperature $T$ is close to the superconducting transition temperature $T_c$, such that $\Delta_{\theta}$ in the denominator of Eqs.~\eqref{F0 for clean system} and~\eqref{eq:anomalous-green-phase-winding} can be neglected. Adding the bulk $F_0$ with impurity-modified $\delta F$ in Eqs.~\eqref{F0 for clean system} and~\eqref{eq:anomalous-green-phase-winding} and neglecting $\Delta_{\theta}$ in denominator, we obtain 
\begin{equation}\label{eq:approx-F-perturbed}
    F(\vec r,\vec r,i\omega_n)\propto \Delta_0-ug(r)\Delta_{\theta_{\vec r}}
\end{equation}
where $\Delta_0=\int \Delta_{\theta} \frac{d\theta}{2\pi}$ is the $s$-wave component of the order parameter, $u={U k_F}/{\pi \hbar {v_F^*}}$ is the dimensionless impurity strength and $g(r)={\cos(2k_Fr)}/{k_Fr}$ describes Friedel-like oscillations. Here, $\theta_{\vec r}$ corresponds to the angle between the $x$ axis and the line connecting the point $\vec r$ to the impurity at origin. Since the coherence length $\xi={\hbar {v_F^*}}/{\Delta}$ is usually very large compared to the Fermi wavelength, we consider the region where $\lambda_F \ll r \ll \xi$ and thus the exponential decay in Eq.~\eqref{eq:anomalous-green-phase-winding} can be neglected. 

Substituting $F(\vec r,\vec r,i\omega_n)$ from Eq.~\eqref{eq:approx-F-perturbed} into ~\eqref{Ic2 for fork tip general},~\eqref{complex F} we obtain the spatial dependence of the critical current,
\begin{equation}\label{eq:fork}
\begin{aligned}
    I^2_{\text{fork-tip}}(\vec R) & \propto |2\Delta_0-ug(\vec r_1)\Delta_{\theta_1}-ug(\vec r_2)\Delta_{\theta_2}|^2
\end{aligned}
\end{equation} 
where $\vec r_1$, $\vec r_2$, $\theta_1$, $\theta_2$ are depicted in Fig.~\ref{fig:fork-tip}(a).


The spatial distribution of the critical current in fork-tip STM for different phase windings is presented in Fig.~\ref{fig:no-noise-plots}. Since our semiclassical formalism is strictly valid for $r\gg \lambda_F$, we only display the pattern where tips are far from the impurity. If there is no phase coherence between two superconducting tips, the phase fluctuations average out the interference term in Eq.~\eqref{two-tip current general form} to zero. The square of total critical current is thus simply a sum of the squares of critical currents through the two tips and thus the overall signal consists of two independent sets of Friedel oscillations. In contrast, when the two tips are phase-coherent, the phase winding of the order parameter generates distinct spatial patterns, from which the angular dependence of the phase can be reconstructed.

\begin{figure}
    \centering
    \includegraphics[width=0.83\linewidth]{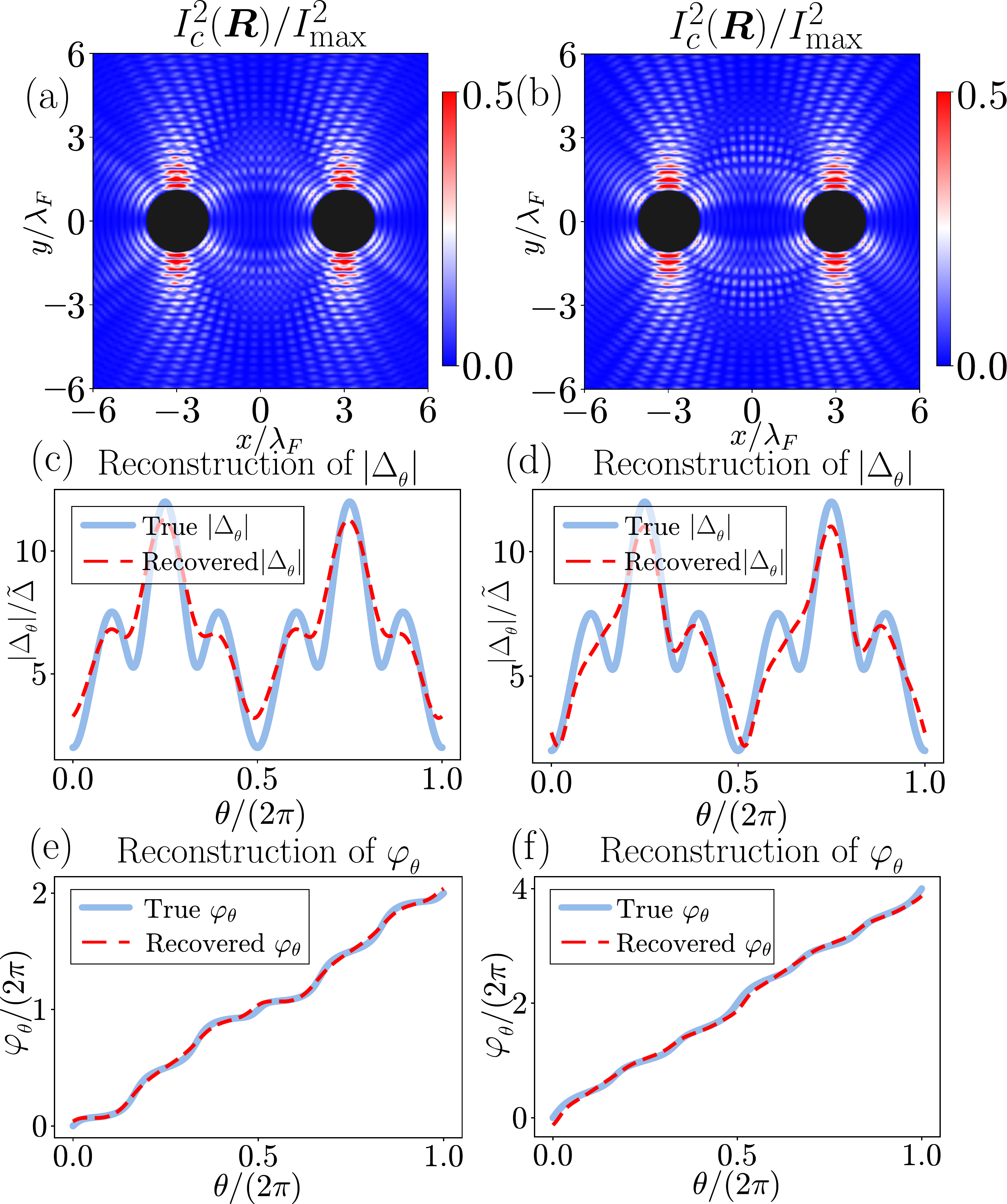}
    \caption{Reconstruction of order parameter from the squared fork-tip critical current $I^2_c$. (a) Spatial map of $I^2_c$ point impurity for order parameter $\Delta_1(\theta) \propto (6 e^{i2\theta} + e^{-i2\theta}-3 e^{i4\theta}- 2e^{-i4\theta})$ defined in Eq.~\eqref{two deltas}. (c,e) Magnitude and phase of $\Delta_1(\theta)$ reconstructed by fitting the signal in panel (a) using angular harmonic expansion in Eq.~\eqref{Delta angular harmonics}. (b) The same as (a) but for $\Delta_2(\theta) \propto (-3 e^{i2\theta}-2 e^{-i2\theta}+6 e^{i4\theta}+ e^{-i4\theta})$ from Eq.~\eqref{two deltas}. (d,f) Corresponding reconstruction of $|\Delta_2(\theta)|$ and $\varphi_2(\theta)$. Although $\Delta_1(\theta)$ and $\Delta_2(\theta)$ have the same magnitude, their phase windings are different. The agreement between reconstructed and original curves demonstrates that both the magnitude and the phase of the order parameter can be recovered from the data. 
    Here we have set identical tunneling amplitude in both tips for simplicity. The robustness of phase recovery for unequal critical currents is demonstrated in Appendix D in SM~\cite{SM_fork-tip}. 
    }
    \label{fig:all_plots}
\end{figure}


Since the phase winding cannot be inferred by visual inspection of the spatial map of fork-tip critical current, we develop a fitting procedure to extract the phase.
If the $s$-wave component of $\Delta_\theta$ is comparable to the chiral components, then the interference term ($\mathcal{O}(u^2)$) in Eq.~\eqref{eq:fork} will be dominated by terms linear in the small parameter $u$.
Therefore, we focus on the case where the $s$-wave component is negligible ($\Delta_0=0$) and expand the order parameter in terms of the angular harmonics,
\begin{equation}\label{Delta angular harmonics}
\Delta_\theta 
=
\Delta_2 e^{i2\theta}
+\Delta_{-2} e^{-i2\theta}
+\Delta_4 e^{i4\theta}
+\Delta_{-4} e^{-i4\theta}
+\cdots
\end{equation}
with arbitrary complex coefficients $\Delta_m$.
The advantage of the fork-tip measurement over a single tip becomes evident when the two order parameters with identical angular profile of magnitude but different phase dependence are considered, 
the simplest examples being $\Delta_\theta=e^{2i\theta}$ and $\Delta_\theta=e^{4i\theta}$. We demonstrate the utility of the proposed method with a more complicated example featuring two gap functions with identical angle-dependence of magnitude but different phase winding,
\begin{align}\label{two deltas}
    \Delta_1(\theta) &= \tilde \Delta (6 e^{i2\theta} + e^{-i2\theta}-3 e^{i4\theta}- 2e^{-i4\theta}), \notag \\
    \Delta_2(\theta) &= \tilde \Delta (-3 e^{i2\theta}-2 e^{-i2\theta}+6 e^{i4\theta}+ e^{-i4\theta}).
\end{align}

The spatial distributions of the squared critical current of the fork-tip for these two order parameters are displayed in Fig.~\ref{fig:all_plots}. 
We use a nonlinear fitting procedure to obtain the relative strengths of $\Delta_2,\Delta_{-2},\Delta_{4},\Delta_{-4}$ (defined in Eq.~\eqref{Delta angular harmonics}), from which the global phase winding as well as the angular dependence of the gap function magnitude can be recovered (up to overall global phase), as demonstrated in Fig.~\ref{fig:all_plots} (see Appendix C~\cite{SM_fork-tip} for details).

\textit{\color{magenta} Discussion and Conclusion---}
We proposed a phase-coherent superconducting fork-tip STM that converts the complex phase structure of an unconventional superconductor into a directly measurable real-space pattern. Near an impurity, 
anomalous Green's function inherits the complex phase of the gap function in momentum space. Interference between the Josephson currents at the two tips provides access to the relative phase difference at two points, and measuring the total critical current allows reconstructing the angle-dependence of both the magnitude and complex phase of the gap $\Delta_{\vec k}$.

Importantly, unlike a single-tip Josephson experiment, the fork-tip setup preserves the relative phase information despite thermal fluctuation-induced phase diffusion, 
as the common fluctuation of their coherent phase drops out of the phase difference. It can therefore distinguish superconductors with identical gap magnitudes but different complex phase windings. Since the fork-tip critical current is invariant under $\Delta_{\theta} \rightarrow \Delta^*_{\theta}$, the phase can be recovered only up to complex conjugation, and opposite chiralities (e.g.\ $d+id$ vs.\ $d-id$) cannot be distinguished. In an experiment, there will be additional periodic modulations due to the Wannier orbitals, which can be Fourier filtered since they are periodic.

Dual and multi-tip superconducting STMs were recently considered both experimentally~\cite{Voigtlander_invited_2018,Nakayama_development_2012, Leeuwenhoek_modeling_2020, Roychowdhury_a_2014, Roychowdhury_development_2014, Liao_simultaneously_2017, Liao_investigation_2019} and theoretically~\cite{Coleman_triplet_2020, Szumniak_local_2026, Ptok_probing_2020}. {In Ref.~\cite{Coleman_triplet_2020}, the authors theoretically proposed to probe periodic spatial modulations of the order parameter with a dual-tip STM. There, the authors considered a system at zero temperature and suggested using a large shunt capacitance to reduce quantum phase fluctuations. In our proposal, the system is in the phase-diffusive limit as $k_B T \gg E_J$, where thermal fluctuations dominate over quantum fluctuations, but the critical current can be obtained even when the charging energy $E_C = e^2/2C$ is larger than the Josephson coupling $E_J$~\cite{Ingold_cooper-pair_1994}. Although thermal fluctuations wash out the phase difference between the coherent tips and the superconductor, the phase difference at the two points on the superconductor can be recovered by measuring the critical Josephson current using the Ivachenko-Zil'berman relation~\cite{Ivanchenko1968, Ambegaokar_voltage_1969, Grabert_single_1992, Ingold_cooper-pair_1994} (see Eq.~\eqref{eq:I-V-relation-current-bias}).}
Earlier experiments on dual-tip STM (Refs.~\cite{Roychowdhury_a_2014, Roychowdhury_development_2014, Liao_simultaneously_2017, Liao_investigation_2019}) focused on detecting the Aharonov-Bohm phase induced by an external magnetic field, requiring the junctions to enclose a large area. In our proposal the relative phase difference is generated intrinsically near an impurity by the chiral superconductor, so the tip separation can be microscopic, potentially as small as several Fermi wavelengths. The smaller separation substantially reduces the length scale over which the phase coherence must be maintained, which may make a phase-coherent fork-tip experimentally accessible. 

The fork-tip protocol introduced in this work can open up new routes for detecting topological and unconventional superconductivity in twisted bilayer cuprates
~\cite{Zhao_time-reversal_2023, Can_high-temperature_2021}, bilayer nickelates~\cite{Ji_time-reversal_2025}, Sr$_2$RuO$_4$~\cite{Pustogow_constraints_2019}, and rhombohedral~\cite{Zhou_superconductivity_2021, Patterson_superconductivity_2025, Zhang_enhanced_2023, Holleis_nematicity_2025, Li_tunable_2024, Chatterjee_inter-valley_2022, Ghazaryan_unconventional_2021} and moir\'e graphene~\cite{Cao_unconventional_2018, Yankowitz_tuning_2019, Lu_superconductors_2019, Dutta_electric_2025, Oh_evidence_2021, Cao_nematicity_2021, Kim_evidence_2022, Park_experimental_2026, Su_superconductivity_2023}. Here we considered a spin-singlet superconductor probed by an $s$-wave superconducting tip. For candidate spin-triplet superconductors such as UPt$_3$, UTe$_2$~\cite{Ran_nearly_2019, Jiao_chiral_2020, Aoki_unconventional_2022, Joynt_the_2002}, and certain phases of rhombohedral graphene~\cite{Han_signatures_2025, Kalantre_2026, Hua_Multi-Knob_2026, Kumar_Pervasive_2025}, an analogous fork-tip scheme would require a superconducting STM tip with nontrivial pairing, possibly featuring a Majorana bound state~\cite{Kashuba_majorana_2017, Lauke_friedel_2018}. Extending our analysis to this case, as well as to realistic band structures with multi-component order parameters, is left for future work. 
{Moreover, the possibility of formation of a bound odd-frequency Cooper pair between the tip and the impurity~\cite{Balatsky_odd-frequency_2018} was not considered here, and its effects in tunneling are also left for future explorations.
}

More broadly, the fork-tip geometry provides a route towards imaging the angular phase winding of the order parameter, complementary to the existing sign-sensitive and time-reversal-symmetry sensitive probes. It elevates Josephson STM from a local probe of order-parameter amplitude to a local interferometric probe of the angle-dependence of the complex gap function.

\textit{\color{magenta} Code availability---} The codes for reconstructing the angular dependence of the gap from its angular harmonics using nonlinear fitting are available on GitHub~\cite{Panigrahi_codes_fitting_SC_2026}.

\textit{\color{magenta} Acknowledgements---}
We thank Alexander Balatsky, Peter Hirschfeld, Daniel Kaplan, Elio K{\"o}nig, {Wan-Ting} Liao and Khachatur Nazaryan for insightful discussions.

\bibliographystyle{apsrev4-2}   

\begin{thebibliography}{9}

\bibitem{Read_paired_2000}
N. Read, and D. Green, Paired states of fermions in two dimensions with breaking of parity and time-reversal symmetries and the fractional quantum Hall effect, \href{https://doi.org/10.1103/PhysRevB.61.10267}{Phys. Rev. B \textbf{61}, 10267 (2000)}.

\bibitem{Kallin_chiral_2016}
C. Kallin, and J. Berlinsky, Chiral superconductors, \href{https://doi.org/10.1088/0034-4885/79/5/054502}{Rep. Prog. Phys. \textbf{79}, 054502 (2016)}.

\bibitem{Sato_topological_2017}
M. Sato, and Y. Ando, Topological superconductors: a review, \href{https://doi.org/10.1088/1361-6633/aa6ac7}{Rep. Prog. Phys. \textbf{80}, 076501 (2017)}.

\bibitem{Damascelli_angle-resolved_2003}
A. Damascelli, Z. Hussain, and Z. X. Shen, Angle-resolved photoemission studies of the cuprate superconductors, \href{https://doi.org/10.1103/RevModPhys.75.473}{Rev. Mod. Phys. \textbf{75}, 473 (2003)}.

\bibitem{Matsuda_nodal_2006}
Y. Matsuda, K. Izawa, and I. Vekhter, Nodal structure of unconventional superconductors probed by angle resolved thermal transport measurements, \href{https://doi.org/10.1088/0953-8984/18/44/R01}{J. Phys.: Condens. Matter \textbf{18}, R705--R752 (2006)}.

\bibitem{Pustogow_constraints_2019}
A. Pustogow, Y. Luo, A. Chronister, Y. S. Su, D. A. Sokolov, F. Jerzembeck, A. P. Mackenzie, C. W. Hicks, N. Kikugawa, S. Raghu, E. D. Bauer, and S. E. Brown, Constraints on the superconducting order parameter in $\mathrm{Sr}_{2}\mathrm{Ru}\mathrm{O}_{4}$ from oxygen-17 nuclear magnetic resonance, \href{https://doi.org/10.1038/s41586-019-1596-2}{Nature \textbf{574}, 72--75 (2019)}.

\bibitem{Wollman_experimental_1993}
D. A. Wollman, D. J. Van Harlingen, W. C. Lee, D. M. Ginsberg, and A. J. Leggett, Experimental determination of the superconducting pairing state in YBCO from the phase coherence of YBCO-Pb dc SQUIDs, \href{https://doi.org/10.1103/PhysRevLett.71.2134}{Phys. Rev. Lett. \textbf{71}, 2134 (1993)}.

\bibitem{VanHarlingen_phase-sensitive_1995}
D. J. Van Harlingen, Phase-sensitive tests of the symmetry of the pairing state in the high-temperature superconductors---Evidence for $d_{x^{2}-y^{2}}$ symmetry, \href{https://doi.org/10.1103/RevModPhys.67.515}{Rev. Mod. Phys. \textbf{67}, 515 (1995)}.

\bibitem{Tsuei_pairing_2000}
C. C. Tsuei, and J. R. Kirtley, Pairing symmetry in cuprate superconductors, \href{https://doi.org/10.1103/RevModPhys.72.969}{Rev. Mod. Phys. \textbf{72}, 969 (2000)}.

\bibitem{Luke_time-reversal_1998}
G. M. Luke, Y. Fudamoto, K. M. Kojima, M. I. Larkin, J. Merrin, B. Nachumi, Y. J. Uemura, Y. Maeno, Z. Q. Mao, Y. Mori, H. Nakamura, and M. Sigrist, Time-reversal symmetry-breaking superconductivity in $\mathrm{Sr}_{2}\mathrm{Ru}\mathrm{O}_{4}$, \href{https://doi.org/10.1038/29038}{Nature \textbf{394}, 558--561 (1998)}.


\bibitem{Xia_high_2006}
J. Xia, Y. Maeno, P. T. Beyersdorf, M. M. Fejer, and A. Kapitulnik, High Resolution Polar Kerr Effect Measurements of ${\mathrm{Sr}}_{2}{\mathrm{RuO}}_{4}$ : Evidence for Broken Time-Reversal Symmetry in the Superconducting State, \href{https://doi.org/10.1103/PhysRevLett.97.167002}{Phys. Rev. Lett. \textbf{97}, 167002 (2006)}.

\bibitem{Hoffman_imaging_2002}
J. E. Hoffman, K. McElroy, D. H. Lee, K. M. Lang, H. Eisaki, S. Uchida, and J. C. Davis, Imaging Quasiparticle Interference in Bi$_2$Sr$_2$CaCu$_2$O$_{8+\delta}$, \href{https://doi.org/10.1126/science.1072640}{Science \textbf{297}, 1148--1151 (2002)}.

\bibitem{Hanaguri_coherence_2009}
T. Hanaguri, Y. Kohsaka, M. Ono, M. Maltseva, P. Coleman, I. Yamada, M. Azuma, M. Takano, K. Ohishi, and H. Takagi, Coherence Factors in a High-$T_c$ Cuprate Probed by Quasi-Particle Scattering Off Vortices, \href{https://doi.org/10.1126/science.1166138}{Science \textbf{323}, 923--926 (2009)}.

\bibitem{Fischer_scanning_2007}
{\O}. Fischer, M. Kugler, I. Maggio-Aprile, C. Berthod, and C. Renner, Scanning tunneling spectroscopy of high-temperature superconductors, \href{https://doi.org/10.1103/RevModPhys.79.353}{Rev. Mod. Phys. \textbf{79}, 353 (2007)}.

\bibitem{Engstrom_detecting_2025}
L. Engstr\"{o}m, P. Simon, and A. Mesaros, Detecting the topological winding of superconducting nodes via local density of states, \href{https://doi.org/10.1103/PhysRevB.111.134505}{Phys. Rev. B \textbf{111}, 134505 (2025)}.

\bibitem{Balatsky_impurity-induced_2006}
A. V. Balatsky, I. Vekhter, and J. X. Zhu, Impurity-induced states in conventional and unconventional superconductors, \href{https://doi.org/10.1103/revmodphys.78.373}{Rev. Mod. Phys. \textbf{78}, 373 (2006)}.


\bibitem{Salkola_theory_1996}
M. I. Salkola, A. V. Balatsky, and D. J. Scalapino, Theory of Scanning Tunneling Microscopy Probe of Impurity States in a $D$-Wave Superconductor, \href{https://doi.org/10.1103/PhysRevLett.77.1841}{Phys. Rev. Lett. \textbf{77}, 1841 (1996)}.

\bibitem{Pan_imaging_2000}
S. H. Pan, E. W. Hudson, K. M. Lang, H. Eisaki, S. Uchida, and J. C. Davis, Imaging the effects of individual zinc impurity atoms on superconductivity in $\mathrm{Bi}_{2}\mathrm{Sr}_{2}\mathrm{Ca}\mathrm{Cu}_{2}\mathrm{O}_{8+\delta}$, \href{https://doi.org/10.1038/35001534}{Nature \textbf{403}, 746--750 (2000)}.

\bibitem{Hudson_interplay_2001}
E. W. Hudson, K. M. Lang, V. Madhavan, S. H. Pan, H. Eisaki, S. Uchida, and J. C. Davis, Interplay of magnetism and high-Tc superconductivity at individual Ni impurity atoms in $\mathrm{Bi}_{2}\mathrm{Sr}_{2}\mathrm{Ca}\mathrm{Cu}_{2}\mathrm{O}_{8+\delta}$, \href{https://doi.org/10.1038/35082019}{Nature \textbf{411}, 920--924 (2001)}.

\bibitem{Rodrigo_scanning_2008}
J. Rodrigo, V. Crespo, and S. Vieira, Scanning tunneling spectroscopy of the vortex state in $\mathrm{Nb}\mathrm{Se}_{2}$ using a superconducting tip, \href{https://doi.org/10.1016/j.physc.2007.11.019}{Physica C: Superconductivity and its Applications \textbf{468}, 547--551 (2008)}.

\bibitem{Hamidian_detection_2016}
M. H. Hamidian, S. D. Edkins, S. H. Joo, A. Kostin, H. Eisaki, S. Uchida, M. J. Lawler, E. A. Kim, A. P. Mackenzie, K. Fujita, J. Lee, and J. C. S. Davis, Detection of a Cooper-pair density wave in $\mathrm{Bi}_{2}\mathrm{Sr}_{2}\mathrm{Ca}\mathrm{Cu}_{2}\mathrm{O}_{8+x}$, \href{https://doi.org/10.1038/nature17411}{Nature \textbf{532}, 343--347 (2016)}.

\bibitem{Cho_a_2019}
D. Cho, K. M. Bastiaans, D. Chatzopoulos, G. D. Gu, and M. P. Allan, A strongly inhomogeneous superfluid in an iron-based superconductor, \href{https://doi.org/10.1038/s41586-019-1408-8}{Nature \textbf{571}, 541--545 (2019)}.

\bibitem{Randeria_scanning_2016}
M. T. Randeria, B. E. Feldman, I. K. Drozdov, and A. Yazdani, Scanning Josephson spectroscopy on the atomic scale, \href{https://doi.org/10.1103/PhysRevB.93.161115}{Phys. Rev. B \textbf{93}, 161115 (2016)}.




\bibitem{Smakov_josephson_2001}
J. \v{S}makov, I. Martin, and A. V. Balatsky, Josephson scanning tunneling microscopy, \href{https://doi.org/10.1103/PhysRevB.64.212506}{Phys. Rev. B \textbf{64}, 212506 (2001)}.


\bibitem{Kimura_josephson_2009}
H. Kimura, R. P. Barber, S. Ono, Y. Ando, and R. C. Dynes, Josephson scanning tunneling microscopy: A local and direct probe of the superconducting order parameter, \href{https://doi.org/10.1103/PhysRevB.80.144506}{Phys. Rev. B \textbf{80}, 144506 (2009)}.

\bibitem{Kimura_scanning_2008}
H. Kimura, R. P. Barber, S. Ono, Y. Ando, and R. C. Dynes, Scanning Josephson Tunneling Microscopy of Single-Crystal ${\mathrm{Bi}}_{2}{\mathrm{Sr}}_{2}{\mathrm{CaCu}}_{2}{\mathrm{O}}_{8+\delta}$ with a Conventional Superconducting Tip, \href{https://doi.org/10.1103/PhysRevLett.101.037002}{Phys. Rev. Lett. \textbf{101}, 037002 (2008)}.

\bibitem{Bergeal_mapping_2008}
N. Bergeal, Y. Noat, T. Cren, T. Proslier, V. Dubost, F. Debontridder, A. Zimmers, D. Roditchev, W. Sacks, and J. Marcus, Mapping the superconducting condensate surrounding a vortex in superconducting ${\text{V}}_{3}\text{Si}$ using a superconducting ${\text{MgB}}_{2}$ tip in a scanning tunneling microscope, \href{https://doi.org/10.1103/PhysRevB.78.140507}{Phys. Rev. B \textbf{78}, 140507 (2008)}.

\bibitem{Graham_imaging_2017}
M. Graham, and D. K. Morr, Imaging the spatial form of a superconducting order parameter via Josephson scanning tunneling spectroscopy, \href{https://doi.org/10.1103/PhysRevB.96.184501}{Phys. Rev. B \textbf{96}, 184501 (2017)}.

\bibitem{Moreno_robust_2026}
J. A. Moreno, P. G. Talavera, E. Herrera, S. L. Valle, Z. Li, L. L. Wang, S. Bud’ko, A. I. Buzdin, I. Guillam\'{o}n, P. C. Canfield, and H. Suderow, Robust Two-Dimensional Surface Superconductivity and Vortex Lattice in the Weyl Semimetal $\gamma\text{-}{\mathrm{PtBi}}_{2}$, \href{https://doi.org/10.1103/9cyw-m5zr}{Phys. Rev. Lett. \textbf{137}, 086001 (2026)}.


\bibitem{Naaman_fluctuation_2001}
O. Naaman, W. Teizer, and R. C. Dynes, Fluctuation Dominated Josephson Tunneling with a Scanning Tunneling Microscope, \href{https://doi.org/10.1103/PhysRevLett.87.097004}{Phys. Rev. Lett. \textbf{87}, 097004 (2001)}.

\bibitem{Jack_critical_2016}
B. J\"{a}ck, M. Eltschka, M. Assig, M. Etzkorn, C. R. Ast, and K. Kern, Critical Josephson current in the dynamical Coulomb blockade regime, \href{https://doi.org/10.1103/PhysRevB.93.020504}{Phys. Rev. B \textbf{93}, 020504 (2016)}.


\bibitem{Ast_sensing_2016}
C. R. Ast, B. J\"{a}ck, J. Senkpiel, M. Eltschka, M. Etzkorn, J. Ankerhold, and K. Kern, Sensing the quantum limit in scanning tunnelling spectroscopy, \href{https://doi.org/10.1038/ncomms13009}{Nat. Commun. \textbf{7}, 13009 (2016)}.


\bibitem{AGD}
A. A. Abrikosov, L. P. Gorkov, I. E. Dzyaloshinski, \textit{Methods of Quantum Field Theory in Statistical Physics}
(Dover, 1975).

\bibitem{Nakayama_development_2012}
T. Nakayama, O. Kubo, Y. Shingaya, S. Higuchi, T. Hasegawa, C. Jiang, T. Okuda, Y. Kuwahara, K. Takami, and M. Aono, Development and Application of Multiple‐Probe Scanning Probe Microscopes, \href{https://doi.org/10.1002/adma.201200257}{Advanced Materials \textbf{24}, 1675--1692 (2012)}.

\bibitem{Voigtlander_invited_2018}
B. Voigtl\"{a}nder, V. Cherepanov, S. Korte, A. Leis, D. Cuma, S. Just, and F. L\"{u}pke, Invited Review Article: Multi-tip scanning tunneling microscopy: Experimental techniques and data analysis, \href{https://doi.org/10.1063/1.5042346}{Review of Scientific Instruments \textbf{89}, 101101 (2018)}.

\bibitem{Leeuwenhoek_modeling_2020}
M. Leeuwenhoek, S. Gr\"{o}blacher, M. P. Allan, and Y. M. Blanter, Modeling Green’s function measurements with two-tip scanning tunneling microscopy, \href{https://doi.org/10.1103/PhysRevB.102.115416}{Phys. Rev. B \textbf{102}, 115416 (2020)}.

\bibitem{Roychowdhury_a_2014}
A. Roychowdhury, M. A. Gubrud, R. Dana, J. R. Anderson, C. J. Lobb, F. C. Wellstood, and M. Dreyer, A 30 m{K}, 13.5 T scanning tunneling microscope with two independent tips, \href{https://doi.org/10.1063/1.4871056}{Review of Scientific Instruments \textbf{85}, 043706 (2014)}.

\bibitem{Roychowdhury_development_2014}
A. Roychowdhury, Development of a dual-tip millikelvin Josephson scanning tunneling microscope, Digital Repository at the University of Maryland, http://hdl.handle.net/1903/15756 (2014).

\bibitem{Liao_simultaneously_2017}
W.-T. Liao, C. J. Lobb, F. C. Wellstood, and M. Dreyer, Simultaneously scanning two connected tips in a scanning tunneling microscope, \href{https://doi.org/10.1063/1.4984626}{Journal of Applied Physics \textbf{121}, 214501 (2017)}.


\bibitem{Liao_investigation_2019}
W.-T. Liao, INVESTIGATION OF TUNNELING IN SUPERCONDUCTORS USING A MILLIKELVIN SCANNING TUNNELING MICROSCOPE, Digital Repository at the University of Maryland, http://drum.lib.umd.edu/handle/1903/22117 (2019).


\bibitem{Szumniak_local_2026}
P. Szumniak, D. Loss, and J. Klinovaja, Local and nonlocal STM transport signatures of spin polarization in second order topological superconductors, arXiv preprint {arXiv:2606.26992} (2026).

\bibitem{Coleman_triplet_2020}
P. Coleman, Y. Komijani, and E. J. K\"{o}nig, Triplet Resonating Valence Bond State and Superconductivity in Hund’s Metals, \href{https://doi.org/10.1103/PhysRevLett.125.077001}{Phys. Rev. Lett. \textbf{125}, 077001 (2020)}.

\bibitem{Ptok_probing_2020}
A. Ptok, D. J. Alspaugh, S. G{\l}odzik, A. Kobia{\l}ka, A. M. Ole\'{s}, P. Simon, and P. Piekarz, Probing the chirality of one-dimensional Majorana edge states around a two-dimensional nanoflake in a superconductor, \href{https://doi.org/10.1103/PhysRevB.102.245405}{Phys. Rev. B \textbf{102}, 245405 (2020)}.


\bibitem{Rog_tapping-mode_2026}
M. Rog, T. J. Blom, D. B. Boltje, J. D. de Haan, R. Fermin, J. Niu, Y. C. Doedes, M. P. Allan, and K. Lahabi, Tapping-Mode SQUID-on-Tip Microscopy with Proximity Josephson Junctions, \href{https://doi.org/10.1021/acs.nanolett.5c04571}{Nano Letters \textbf{26}, 1608--1615 (2026)}.

\bibitem{Ding_hyperbolic_2023}
P. Ding, T. Schwemmer, C. H. Lee, X. Wu, and R. Thomale, Hyperbolic Fringe Signal for Twin Impurity Quasiparticle Interference, \href{https://doi.org/10.1103/PhysRevLett.130.256001}{Phys. Rev. Lett. \textbf{130}, 256001 (2023)}.

\bibitem{Panigrahi_Poliakov_tomographic_2026}
A. Panigrahi, V. Poliakov, and L. Levitov, Tomographic imaging of superconducting order using particle–hole interference, \href{https://doi.org/10.1073/pnas.2534730123}{Proceedings of the National Academy of Sciences \textbf{123}, e2534730123 (2026)}.

\bibitem{Panigrahi_Poliakov_particle_hole_2026}
A. Panigrahi, V. Poliakov, and L. Levitov, Particle-Hole Ghost Interference in Superconductors, arXiv preprint {arXiv:2606.06437} (2026).

\bibitem{Lauke_friedel_2018}
L. Lauke, M. S. Scheurer, A. Poenicke, and J. Schmalian, Friedel oscillations and Majorana zero modes in inhomogeneous superconductors, \href{https://doi.org/10.1103/PhysRevB.98.134502}{Phys. Rev. B \textbf{98}, 134502 (2018)}.

\bibitem{Kashuba_majorana_2017}
O. Kashuba, B. Sothmann, P. Burset, and B. Trauzettel, Majorana STM as a perfect detector of odd-frequency superconductivity, \href{https://doi.org/10.1103/PhysRevB.95.174516}{Phys. Rev. B \textbf{95}, 174516 (2017)}.


\bibitem{Larkin1967tunnel}
A.~I. Larkin and Yu.~N. Ovchinnikov,
{Tunnel effect between superconductors in an alternating field},
Zh. Eksp. Teor. Fiz. \textbf{51}, 1535 (1966)
[\href{https://jetp.ras.ru/cgi-bin/e/index/e/24/5/p1035?a=list}{Sov. Phys. JETP \textbf{24}, 1035 (1967)}].

\bibitem{Larkin1967boundary}
A.~I. Larkin, Yu.~N. Ovchinnikov, and M.~A. Fedorov,
{Boundary conditions of the Josephson effect},
Zh. Eksp. Teor. Fiz. \textbf{51}, 683 (1966)
[\href{http://jetp.ras.ru/cgi-bin/e/index/e/24/2/p452?a=list}{Sov. Phys. JETP \textbf{24}, 452 (1967)}].

\bibitem{SM_fork-tip}
See Supplemental Material at XXX-XXXX for derivation of Josephson current for arbitrary spin and momentum dependence of the order parameter, derivation of impurity-induced Green's function, details for the nonlinear fitting procedure, and robustness of the fitting procedure for noisy data and different coupling strengths of the two tips. 

\bibitem{Ivanchenko1968}
Yu.~M.~Ivanchenko and L.~A.~Zil'berman,
The Josephson effect in small tunnel contacts,
Zh.~Eksp.~Teor.~Fiz.~\textbf{55}, 2395 (1968)
[\href{http://jetp.ras.ru/cgi-bin/e/index/e/28/6/p1272?a=list}{Sov.~Phys.~JETP~\textbf{28}, 1272 (1969)}].

\bibitem{Ambegaokar_voltage_1969}
V. Ambegaokar, and B. I. Halperin, Voltage Due to Thermal Noise in the dc Josephson Effect, \href{https://doi.org/10.1103/PhysRevLett.22.1364}{Phys. Rev. Lett. \textbf{22}, 1364 (1969)}.

\bibitem{Grabert_single_1992}
H. Grabert and M. H. Devoret, Single Charge Tunneling, NATO ASI Ser. B, Vol. 294 (Plenum, New York, 1992).

\bibitem{Ingold_cooper-pair_1994}
G. L. Ingold, H. Grabert, and U. Eberhardt, Cooper-pair current through ultrasmall Josephson junctions, \href{https://doi.org/10.1103/PhysRevB.50.395}{Phys. Rev. B \textbf{50}, 395 (1994)}.










\bibitem{Zhao_time-reversal_2023}
S. Y. F. Zhao, X. Cui, P. A. Volkov, H. Yoo, S. Lee, J. A. Gardener, A. J. Akey, R. Engelke, Y. Ronen, R. Zhong, G. Gu, S. Plugge, T. Tummuru, M. Kim, M. Franz, J. H. Pixley, N. Poccia, and P. Kim, Time-reversal symmetry breaking superconductivity between twisted cuprate superconductors, \href{https://doi.org/10.1126/science.abl8371}{Science \textbf{382}, 1422--1427 (2023)}.

\bibitem{Can_high-temperature_2021}
O. Can, T. Tummuru, R. P. Day, I. Elfimov, A. Damascelli, and M. Franz, High-temperature topological superconductivity in twisted double-layer copper oxides, \href{https://doi.org/10.1038/s41567-020-01142-7}{Nat. Phys. \textbf{17}, 519--524 (2021)}.

\bibitem{Ji_time-reversal_2025}
H. Ji, Z. Xie, Y. Chen, G. Zhou, L. Pan, H. Wang, H. Huang, J. Ge, Y. Liu, G. M. Zhang, Z. Wang, Q. K. Xue, Z. Chen, and J. Wang, Time-reversal symmetry breaking superconductivity with electronic glass in nickelate (La, Pr, Sm)$_3$$\mathrm{Ni}_{2}\mathrm{O}_{7}$ films, arXiv preprint {arXiv:2508.16412} (2025).


\bibitem{Zhou_superconductivity_2021}
H. Zhou, T. Xie, T. Taniguchi, K. Watanabe, and A. F. Young, Superconductivity in rhombohedral trilayer graphene, \href{https://doi.org/10.1038/s41586-021-03926-0}{Nature \textbf{598}, 434--438 (2021)}.

\bibitem{Patterson_superconductivity_2025}
C. L. Patterson, O. I. Sheekey, T. B. Arp, L. F. W. Holleis, J. M. Koh, Y. Choi, T. Xie, S. Xu, Y. Guo, H. Stoyanov, E. Redekop, C. Zhang, G. Babikyan, D. Gong, H. Zhou, X. Cheng, T. Taniguchi, K. Watanabe, M. E. Huber, C. Jin, \'{E}. Lantagne-Hurtubise, J. Alicea, and A. F. Young, Superconductivity and spin canting in spin–orbit-coupled trilayer graphene, \href{https://doi.org/10.1038/s41586-025-08863-w}{Nature \textbf{641}, 632--638 (2025)}.

\bibitem{Zhang_enhanced_2023}
Y. Zhang, R. Polski, A. Thomson, \'{E}. Lantagne-Hurtubise, C. Lewandowski, H. Zhou, K. Watanabe, T. Taniguchi, J. Alicea, and S. Nadj-Perge, Enhanced superconductivity in spin–orbit proximitized bilayer graphene, \href{https://doi.org/10.1038/s41586-022-05446-x}{Nature \textbf{613}, 268--273 (2023)}.

\bibitem{Holleis_nematicity_2025}
L. Holleis, C. L. Patterson, Y. Zhang, Y. Vituri, H. M. Yoo, H. Zhou, T. Taniguchi, K. Watanabe, E. Berg, S. Nadj-Perge, and A. F. Young, Nematicity and orbital depairing in superconducting Bernal bilayer graphene, \href{https://doi.org/10.1038/s41567-024-02776-7}{Nat. Phys. \textbf{21}, 444--450 (2025)}.

\bibitem{Li_tunable_2024}
C. Li, F. Xu, B. Li, J. Li, G. Li, K. Watanabe, T. Taniguchi, B. Tong, J. Shen, L. Lu, J. Jia, F. Wu, X. Liu, and T. Li, Tunable superconductivity in electron- and hole-doped Bernal bilayer graphene, \href{https://doi.org/10.1038/s41586-024-07584-w}{Nature \textbf{631}, 300--306 (2024)}.

\bibitem{Chatterjee_inter-valley_2022}
S. Chatterjee, T. Wang, E. Berg, and M. P. Zaletel, Inter-valley coherent order and isospin fluctuation mediated superconductivity in rhombohedral trilayer graphene, \href{https://doi.org/10.1038/s41467-022-33561-w}{Nat. Commun. \textbf{13}, 6013 (2022)}.

\bibitem{Ghazaryan_unconventional_2021}
A. Ghazaryan, T. Holder, M. Serbyn, and E. Berg, Unconventional Superconductivity in Systems with Annular Fermi Surfaces: Application to Rhombohedral Trilayer Graphene, \href{https://doi.org/10.1103/PhysRevLett.127.247001}{Phys. Rev. Lett. \textbf{127}, 247001 (2021)}.

\bibitem{Cao_unconventional_2018}
Y. Cao, V. Fatemi, S. Fang, K. Watanabe, T. Taniguchi, E. Kaxiras, and P. Jarillo-Herrero, Unconventional superconductivity in magic-angle graphene superlattices, \href{https://doi.org/10.1038/nature26160}{Nature \textbf{556}, 43--50 (2018)}.

\bibitem{Yankowitz_tuning_2019}
M. Yankowitz, S. Chen, H. Polshyn, Y. Zhang, K. Watanabe, T. Taniguchi, D. Graf, A. F. Young, and C. R. Dean, Tuning superconductivity in twisted bilayer graphene, \href{https://doi.org/10.1126/science.aav1910}{Science \textbf{363}, 1059--1064 (2019)}.

\bibitem{Lu_superconductors_2019}
X. Lu, P. Stepanov, W. Yang, M. Xie, M. A. Aamir, I. Das, C. Urgell, K. Watanabe, T. Taniguchi, G. Zhang, A. Bachtold, A. H. MacDonald, and D. K. Efetov, Superconductors, orbital magnets and correlated states in magic-angle bilayer graphene, \href{https://doi.org/10.1038/s41586-019-1695-0}{Nature \textbf{574}, 653--657 (2019)}.

\bibitem{Dutta_electric_2025}
R. Dutta, A. Ghosh, S. Mandal, K. Watanabe, T. Taniguchi, H. R. Krishnamurthy, S. Banerjee, M. Jain, and A. Das, Electric Field-Tunable Superconductivity with Competing Orders in Twisted Bilayer Graphene near the Magic Angle, \href{https://doi.org/10.1021/acsnano.4c12770}{ACS Nano \textbf{19}, 5353--5362 (2025)}.

\bibitem{Oh_evidence_2021}
M. Oh, K. P. Nuckolls, D. Wong, R. L. Lee, X. Liu, K. Watanabe, T. Taniguchi, and A. Yazdani, Evidence for unconventional superconductivity in twisted bilayer graphene, \href{https://doi.org/10.1038/s41586-021-04121-x}{Nature \textbf{600}, 240--245 (2021)}.

\bibitem{Cao_nematicity_2021}
Y. Cao, D. Rodan-Legrain, J. M. Park, N. F. Q. Yuan, K. Watanabe, T. Taniguchi, R. M. Fernandes, L. Fu, and P. Jarillo-Herrero, Nematicity and competing orders in superconducting magic-angle graphene, \href{https://doi.org/10.1126/science.abc2836}{Science \textbf{372}, 264--271 (2021)}.

\bibitem{Kim_evidence_2022}
H. Kim, Y. Choi, C. Lewandowski, A. Thomson, Y. Zhang, R. Polski, K. Watanabe, T. Taniguchi, J. Alicea, and S. Nadj-Perge, Evidence for unconventional superconductivity in twisted trilayer graphene, \href{https://doi.org/10.1038/s41586-022-04715-z}{Nature \textbf{606}, 494--500 (2022)}.

\bibitem{Park_experimental_2026}
J. M. Park, S. Sun, K. Watanabe, T. Taniguchi, and P. Jarillo-Herrero, Experimental evidence for nodal superconducting gap in moir\'e graphene, \href{https://doi.org/10.1126/science.adv8376}{Science \textbf{391}, 79--83 (2026)}.

\bibitem{Su_superconductivity_2023}
R. Su, M. Kuiri, K. Watanabe, T. Taniguchi, and J. Folk, Superconductivity in twisted double bilayer graphene stabilized by $\mathrm{W}\mathrm{Se}_{2}$, \href{https://doi.org/10.1038/s41563-023-01653-7}{Nat. Mater. \textbf{22}, 1332--1337 (2023)}.


\bibitem{Han_signatures_2025}
T. Han, Z. Lu, Z. Hadjri, L. Shi, Z. Wu, W. Xu, Y. Yao, A. A. Cotten, O. Sharifi Sedeh, H. Weldeyesus, J. Yang, J. Seo, S. Ye, M. Zhou, H. Liu, G. Shi, Z. Hua, K. Watanabe, T. Taniguchi, P. Xiong, D. M. Zumb\"{u}hl, L. Fu, and L. Ju, Signatures of chiral superconductivity in rhombohedral graphene, \href{https://doi.org/10.1038/s41586-025-09169-7}{Nature \textbf{643}, 654--661 (2025)}.

\bibitem{Kalantre_2026}
S. S. Kalantre, B. H. Alexander, J. May-Mann, J. Herzog-Arbeitman, M. Hocking, Q. Cao, K. Watanabe, T. Taniguchi, D. Goldhaber-Gordon, A. J. Mannix, T. Devakul, Y. H. Kwan, D. E. Parker, A. Sharpe,
Fermiology and the Candidate Chiral Superconductor in Rhombohedral Tetralayer Graphene, arXiv preprint {arXiv:2606.05356} (2026).

\bibitem{Hua_Multi-Knob_2026}
Z. Hua, S. Ye, P. Pattanakanvijit, G. Shi, T. Han, E. Aitken, J. Yang, J. Seo, H. Liu, R. Hao, K. Xiao, J. Guo, V. T. Phong, K. Watanabe, T. Taniguchi, C. Huang, C. Lewandowski, L. Ju, P. Xiong, Z. Lu,
Multi-Knob Switchable Chiral Superconductivity Quartet in Rhombohedral Graphene, arXiv preprint {arXiv:2607.06520} (2026).

\bibitem{Kumar_Pervasive_2025}
M. Kumar, D. Waleffe, A. Okounkova, R. Tejani, K. Watanabe, T. Taniguchi, \'E. Lantagne-Hurtubise, J. Folk, M. Yankowitz,
Pervasive spin-triplet superconductivity in rhombohedral graphene, arXiv preprint {arXiv:2511.16578} (2025).



\bibitem{Joynt_the_2002}
R. Joynt, and L. Taillefer, The superconducting phases of ${\mathrm{UPt}}_{3}$, \href{https://doi.org/10.1103/RevModPhys.74.235}{Rev. Mod. Phys. \textbf{74}, 235 (2002)}.

\bibitem{Ran_nearly_2019}
S. Ran, C. Eckberg, Q. P. Ding, Y. Furukawa, T. Metz, S. R. Saha, I. L. Liu, M. Zic, H. Kim, J. Paglione, and N. P. Butch, Nearly ferromagnetic spin-triplet superconductivity, \href{https://doi.org/10.1126/science.aav8645}{Science \textbf{365}, 684--687 (2019)}.

\bibitem{Jiao_chiral_2020}
L. Jiao, S. Howard, S. Ran, Z. Wang, J. O. Rodriguez, M. Sigrist, Z. Wang, N. P. Butch, and V. Madhavan, Chiral superconductivity in heavy-fermion metal $\mathrm{U}\mathrm{Te}_{2}$, \href{https://doi.org/10.1038/s41586-020-2122-2}{Nature \textbf{579}, 523--527 (2020)}.

\bibitem{Aoki_unconventional_2022}
D. Aoki, J. P. Brison, J. Flouquet, K. Ishida, G. Knebel, Y. Tokunaga, and Y. Yanase, Unconventional superconductivity in UTe\textsubscript{2}, \href{https://doi.org/10.1088/1361-648x/ac5863}{J. Phys.: Condens. Matter \textbf{34}, 243002 (2022)}.

\bibitem{Balatsky_odd-frequency_2018}
A. V. \mbox{Balatsky}, S. S. \mbox{Pershoguba}, and C. \mbox{Triola}, Odd-frequency Pairing in Conventional Josephson Junctions, \href{https://arxiv.org/abs/1804.07244}{arXiv:1804.07244} (2018).


\bibitem{Panigrahi_codes_fitting_SC_2026}
A. Panigrahi, \href{https://github.com/archisman-panigrahi/fork-tip_Josephson_fitting}{Codes for reconstructing the angular dependence of gap function with fork-tip STM}, GitHub (2026).

\end{thebibliography}

\end{document}


\title{\bf Supplemental Material:\\ 
Imaging phase winding in topological superconductors with a fork-tip Josephson STM}

\author{Vladislav Poliakov}
\email{vlad\_p@mit.edu}
\affiliation{Department of Physics, Massachusetts Institute of Technology, Cambridge, Massachusetts 02139, USA}

\author{Archisman Panigrahi}
\email{archi137@mit.edu}
\affiliation{Department of Physics, Massachusetts Institute of Technology, Cambridge, Massachusetts 02139, USA}
\date{\today}
\begin{abstract}
The Supplemental Material contains A) the derivation of Josephson current for arbitrary spin and momentum-dependence of the order parameter, B) derivation of impurity-induced Green's function, C) details for the non-linear fitting procedure, and D) robustness of the fitting procedure for noisy data and different coupling strengths of the two tips.
\end{abstract}

\maketitle
\vspace{-1cm}
\appendix

\section{Josephson current for arbitrary spin and momentum-dependent order parameter}\label{appendix:Josephson current}
We derive Josephson current between a two-dimensional system coupled to a superconducting STM tip in the weak coupling limit. The  total Hamiltonian of the system is
\begin{eqnarray}  \hat{H}=\hat{H}_{\text{tip}}+\hat{H}_{\text{sys}}+\hat{H}_{\text{t}}
\end{eqnarray}
where $\hat{H}_{\text{tip}}$ describes the electronic energy levels in the tip, $\hat{H}_{\text{sys}}$ describes the energy levels in the superconducting system, and $\hat{H}_{\text{t}}$ is the tunneling Hamiltonian.
We assume a local channel independent tunneling amplitude between the tip at position $\vec { r_0}$ and the system,
\begin{eqnarray}
    \hat{H}_t=\sum_\nu \Big[t c^\dagger_{{\vec r_0}}d_\nu +t^* d^\dagger_\nu c_{{\vec r_0}} \Big]
\end{eqnarray}
where $c_{{\vec r_0}}$ is an electronic annihilation operator in the system at position $\vec { r_0}$, and $d_\nu$ is an annihilation operator in the tip channel $\nu$. For simplicity spin indices are suppressed below and will be restored later. The current can be found from a Heisenberg's equation for the number of particles inside the material:
\begin{eqnarray}
    \langle I \rangle=-e\left\langle \frac{d}{dt}\sum_{\vec p} c^\dagger_{\vec p} c_{\vec p}\right\rangle=-i e\left\langle \left[H_t,\sum_{\vec p} c^\dagger_{\vec p} c_{\vec p}\right] \right\rangle
\end{eqnarray}
Evaluating this commutator gives an expression for the average current:
\begin{eqnarray}
    \langle I \rangle=-2e \;\text{Im}\Big[t\sum_\nu \left\langle c^\dagger_{{\vec r_0}}d_\nu\right\rangle \Big]
\end{eqnarray}
In the interaction picture the expectation value at finite temperature is found from expanding the evolution operator
\begin{eqnarray}
    \langle c^\dagger_{{\vec r_0}}d_\nu\rangle=\left\langle T_\tau c^\dagger_{{\vec r_0}}(0)d_\nu(0) e^{-\int_0^\beta H_t(\tau)d\tau}\right\rangle\approx -\int_0^\beta d\tau \sum_\mu \left\langle T_\tau c^\dagger_{{\vec r_0}}(0)d_\nu(0) \Big(t c^\dagger_{{\vec r_0}}(\tau)d_\mu(\tau)+t^* d^\dagger_\mu(\tau) c_{{\vec r_0}}(\tau)\Big)\right\rangle
\end{eqnarray}
The second term gives a regular quasiparticle tunneling current. At temperatures smaller than the superconducting gap and zero voltage bias, this current is negligible, so we omit it. The first term, however, produces a Josephson current, which is proportional to anomalous Green's functions:
\begin{eqnarray}
    -t\int_0^\beta d\tau \sum_\mu \langle T_\tau c^\dagger_{{\vec r_0}}(0)d_\nu(0) c^\dagger_{{\vec r_0}}(\tau)d_\mu(\tau)\rangle =t\int_0^\beta d\tau F^\dagger ({\vec r_0}, {\vec r_0},\tau)F_{\text{tip}}(\nu,-\tau)
\end{eqnarray}
For each tip channel $\nu$ the summation over $\mu$ selects only time-reversal partner of channel $\nu$. Fourier transforming to Matsubara frequency then gives:
\begin{eqnarray}
    \langle I \rangle=-2e t^2 T\sum_{i\omega_n}\;\text{Im}\Big[F^\dagger({\vec r_0},  {\vec r_0},i\omega_n) \sum_\nu F_{\text{tip}}(\nu,i\omega_n) \Big].
\end{eqnarray}

Here and below we assume that $t$ is real so it does not produce an additional phase. 
We use the anomalous Green's functions:
\begin{align}
F_{\alpha\beta}(\vec{r}_1,\tau_1;\vec{r}_2,\tau_2)
&=
-
\left\langle
T_\tau\,
\psi_\alpha(\vec{r}_1,\tau_1)
\psi_\beta(\vec{r}_2,\tau_2)
\right\rangle ,\\
F_{\alpha\beta}^{\dagger}(\vec{r}_1,\tau_1;\vec{r}_2,\tau_2)
&=
-
\left\langle
T_\tau\,
\psi_\alpha^{\dagger}(\vec{r}_1,\tau_1)
\psi_\beta^{\dagger}(\vec{r}_2,\tau_2)
\right\rangle.
\end{align}
which obey the relation 
\begin{eqnarray}
    F_{\alpha\beta}^*(\vec{r}_1,\tau_1;\vec{r}_2,\tau_2)=F_{\beta\alpha}^\dagger(\vec{r}_2,\tau_2;\vec{r}_1,\tau_1)
\end{eqnarray}

Alternatively, for complex frequency $z$, if $F(z)$ {is analytic everywhere except for the real axis}, then one can rewrite the summation over Matsubara frequencies as an integral over the real axis:
\begin{equation}
    T\sum_{i\omega_n} \text{Im} \Big[ F^\dagger({\vec r_0}, {\vec r_0} ,i\omega_n) F_{\text{tip}}(\nu,i\omega_n)\Big]= -\text{Im}\int n_F(\omega) \Big[F^\dagger({\vec r_0}, {\vec r_0},\omega+i0) F_{\text{tip}}(\nu,\omega+i0)- F^\dagger({\vec r_0}, {\vec r_0},\omega-i0) F_{\text{tip}}(\nu,\omega-i0)\Big]\frac{d\omega}{2\pi i}.
\end{equation}
Using corresponding retarded and advanced analytical continuations we arrive at the expression
\begin{eqnarray}
    \langle I \rangle = 2et^2 \sum_\nu\int n_F(\omega) \text{Re}\Big[F_R^\dagger({\vec r_0}, {\vec r_0},\omega)F_{R,tip}(\nu,\omega)-F_A^\dagger({\vec r_0}, {\vec r_0},\omega)F_{A,tip}(\nu,\omega) \Big] \frac{d\omega}{2\pi}
\end{eqnarray}
This formula, after including the spin degeneracy factor and detaching the overall phase of the order parameter is equivalent to the one used in  Ref.~\cite{Graham_imaging_2017}.

 In order to verify these results let us calculate the current between two s-wave superconductors. For simplicity let us also assume that the anomalous Green's function of the tip has the same form as $F(r,r,i\omega_n)$ in a 2D superconductor. For an isotropic system, the current takes the form
\begin{eqnarray}
    \langle I \rangle=-2e t^2 \left(\frac{k_F}{2{v_F^*}}\right)^2 T\sum_{i\omega_n}\;\frac{\text{Im} [\Delta^*_{\text{sys}}\Delta_{\text{tip}}]}{\sqrt{\omega_n^2+|\Delta_{\text{sys}}|^2}\sqrt{\omega_n^2+|\Delta_{\text{tip}}|^2}}
\end{eqnarray}
The imaginary part produces a sine of phase difference between two superconductors, leading to a standard expression for Josephson current. 

We finally restore the spin structure and obtain the general expression for tunneling operator:
\begin{eqnarray}
    -t \sum_{ab}\int_0^\beta d\tau \sum_{\nu'} \langle T_\tau c^\dagger_{a}({\vec r_0},0)d_a(\nu,0) c^\dagger_{b}({\vec r_0},\tau)d_b({\nu'},\tau)\rangle =t\sum_{ab}\int_0^\beta d\tau F_{ab}^\dagger ({\vec r_0}, {\vec r_0},\tau)F_{ab}^{\text{tip}}(\nu,-\tau)
\end{eqnarray}
where $a$ and $b$ are spin indices. In case one can separate the spin structure from the Green's function:
\begin{eqnarray}
    F_{ab}(\vec r_1,\vec r_2,\tau)\equiv \Gamma_{ab}F(\vec r_1,\vec r_2,\tau)
\end{eqnarray}
then the spin simply adds a factor in front of the average current,
\begin{eqnarray}\label{eq:current-with-spin-structure}
    \langle I \rangle=-2e t^2 \Big[\sum_{ab}(\Gamma_{ab})^*\Gamma_{ab}^{\text{tip}} \Big] T\sum_{i\omega_n}\;\text{Im}\Big[F^\dagger({\vec r_0}, {\vec r_0},i\omega_n) \sum_\nu F_{\text{tip}}(\nu,i\omega_n) \Big].
\end{eqnarray}
If the spin structure of the Cooper pair is the same both in the system and in an STM tip, then the spin sum simply gives a factor of 2. If the spin structure is orthogonal to each other (for example, singlet and triplet or $|\uparrow \uparrow\rangle$ and $|\downarrow \downarrow\rangle$) then the spin sum is zero, suggesting that there is no Josephson tunneling, because a spin-singlet Cooper pair cannot coherently tunnel to a spin-triplet superconductor.

\section{Computation of impurity-modified anomalous Green's function} \label{appendix: details of Green's functions}
In the semiclassical limit $\vec r \gg \lambda_F$ (also known as Eilenberger limit), the real-space Green's function can be computed with saddle-point method (see supplementary information in Ref.~\cite{Panigrahi_Poliakov_tomographic_2026}),
\begin{equation}
G_0(\vec r,i\omega_n)=\frac{D(\vec r)}{\sqrt{\omega_n^2+|\Delta_{\theta_{\vec r}}|^2}} \times \begin{pmatrix}
        i\omega_n \cos \gamma_r - \sqrt{\omega_n^2+|\Delta_{\theta_{\vec r}}|^2}\sin \gamma_r & \frac{1}{2}\Big[\Delta_{\theta_{\vec r}} e^{i\gamma_r}+\Delta_{{\theta_{\vec r}}+\pi}e^{-i\gamma_r} \Big]\\
        
        \frac{1}{2}\Big[\Delta_{\theta_{\vec r}}^* e^{i\gamma_r}+\Delta_{{\theta_{\vec r}}+\pi}^*e^{-i\gamma_r} \Big] & i\omega_n \cos \gamma_r + \sqrt{\omega_n^2+|\Delta_{\theta_{\vec r}}|^2}\sin \gamma_r
    \end{pmatrix}
\end{equation}
where $D(\vec r)=-\sqrt{\frac{k_F}{2\pi {v_F^*}^2 r}}e^{-\frac{r}{{v_F^*}}\sqrt{\omega_n^2+|\Delta_{\theta_{\vec r}}|^2}}$, $\gamma_r=(k_F r-{\pi}/{4})$ is the dimensionless distance, and ${\theta_{\vec r}} = \tan^{-1}(y/x)$ is an angle of vector $\vec r$ with the $x$-axis. Here, the gap function magnitudes are equal for $\pm \vec k$ pairs, and for both spin-singlet and triplet superconductors, the gap functions at opposite momenta have equal magnitudes, i.e., $|\Delta_{{\theta_{\vec r}}}|=|\Delta_{{\theta_{\vec r}} + \pi}|$.

For an impurity placed at $\vec r=0$, we obtain the impurity modified Green's function $\delta G = G_0 (U \tau_z) G_0$, leading to the following form of impurity-modified anomalous Green's function $\delta F(\vec r, \vec r,i\omega_n) = \text{Tr}_{\tau}\left[\tau^{-} \delta G(\vec r, \vec r,i\omega_n)\right]$,
\begin{eqnarray}
    \delta F(\vec r, \vec r,i\omega_n)=U \frac{k_F}{2\pi {v_F^*}^2}\frac{\sin(2\gamma_r)e^{-\frac{2r}{{v_F^*}}\sqrt{\omega_n^2+|\Delta_{\theta_{\vec r}}|^2}}}{r}
    \times 
\begin{cases}
\frac{-|\Delta_{\theta_{\vec r}}| e^{i\varphi_{\theta_{\vec r}}}}{\sqrt{\omega_n^2+|\Delta_{\theta_{\vec r}}|^2}}, & \Delta_{\vec k}=\Delta_{-\vec k} (\rm{spin-singlet}),\\
\frac{\omega_n |\Delta_{\theta_{\vec r}}| e^{i\varphi_{\theta_{\vec r}}}}{\omega_n^2+|\Delta_{{\theta_{\vec r}}}|^2}, & \Delta_{\vec k}=-\Delta_{-\vec k} (\rm{spin-triplet}).
\end{cases}
\end{eqnarray}

From equation Eq.~\eqref{eq:current-with-spin-structure} we obtain,
\begin{eqnarray}
    I \propto t^2 T \sum_{i\omega_n} \text{Im}\Big[\delta F^*(\vec r_0, \vec r_0,i\omega_n)\frac{\Delta_{\rm tip}}{\sqrt{\omega_n^2+|\Delta_{\rm tip}|^2}} \Big].
\end{eqnarray}
Since the only complex part in the anomalous Green's function is the order parameter in the numerator, we can immediately write,
\begin{eqnarray}
    I(\vec r_0) =I_c(\vec r_0)\sin(\varphi_{\theta_{\vec r_0}}-\varphi_{\rm tip})
\end{eqnarray}
where $\varphi_{\rm tip}$ is an overall phase of the order parameter in the tip. For spin-singlet superconductors, the critical current $I_c(\vec r)$ is given by,
\begin{equation}
    I_c(\vec r_0)\propto t^2 \frac{\sin(2\gamma_{r_0})}{k_Fr_0}
T\sum_{i\omega_n}\frac{-|\Delta_{\theta_{\vec r_0}}| }{\sqrt{\omega_n^2+|\Delta_{\theta_{\vec r_0}}|^2}}\frac{|\Delta_{\rm tip}|}{\sqrt{\omega_n^2+|\Delta_{\rm tip}|^2}}.
\end{equation}

Note: Here $\vec r_0$ corresponds to the position of a single tip, whereas the quantity $\vec R$ corresponds to the midpoint of a fork tip, which has two individual tips at $\vec r_{1,2} = \vec R \pm \frac{\vec a}{2}$.

\section{Fitting procedure to reconstruct momentum-dependence of phase}\label{app:curve-fitting}

We fit the generated signal of squared critical current (that may or may not have noise) with the function 
\begin{equation}
I_c^2(\vec r) \propto |\mathcal F(\vec r_1) + \mathcal F(\vec r_2)|^2,
\end{equation}
where $\mathcal F_i$ is of the form,
\begin{equation}
\mathcal F_i = - u g(r_i) m(r_i) \Delta_{\theta_{{\vec r}_i}},
\end{equation}
where $g(r) = \cos(2 k_F r)/(k_F r)$ and $m(r)$ is a function masking the signal at small distances, where the Eilenberger approximation is not valid. Here, for simplicity, we have set the $s$-wave component $\Delta_0 = 0$.

We minimize the cost function
\begin{equation}
    S = \sum_{x,y} \left||\mathcal F(\vec r_1) + \mathcal F(\vec r_2)|^2 - {I_c^2}_{,\rm data} \right|^2 
\end{equation}
with respect to the unknown parameters $ \Delta_{2}, \Delta_{-2}, \Delta_{4}, \Delta_{-4}, \Delta_{6}, \Delta_{-6}$, which are the angular harmonics in the gap function, $\Delta_\theta 
=
\Delta_2 e^{i2\theta}
+\Delta_{-2} e^{-i2\theta}
+\Delta_4 e^{i4\theta}
+\Delta_{-4} e^{-i4\theta}
+\Delta_6 e^{i6\theta}
+\Delta_{-6} e^{-i6\theta} $.

The gauge transformation $\Delta_\theta \rightarrow \Delta_\theta e^{i\chi}$, which is transformation of overall phase, does not alter the observed spatial pattern, and a naive minimization will result in some arbitrary overall phase.
In order to make the results of the fitting procedure reproducible, we adopt the convention that the coefficient $\Delta_m$ of the leading Fourier harmonic is a real positive number. Moreover, the transformation $\Delta_\theta \rightarrow \Delta_\theta^*$ leaves the observed pattern invariant, the we adopt a convention that the code always chooses the positive chirality.

The codes are written in the Julia programming language, and they use the \texttt{Optim.optimize} function for non-linear fitting. Also, it begins the optimization in different families of the fitted parameters, where the individual angular harmonics are chosen to be real. The code is available on GitHub~\cite{Panigrahi_codes_fitting_SC_2026}.

The code also works for unequal tunneling amplitudes at the two tips, described in detail in Appendix~\ref{app:sec:unequal-tip}.

\section{Robustness of the fitting procedure against different tunneling strength of the two tips}~\label{app:sec:unequal-tip}

\begin{figure}[ht]
    \centering
    \includegraphics[width=1.0\linewidth]{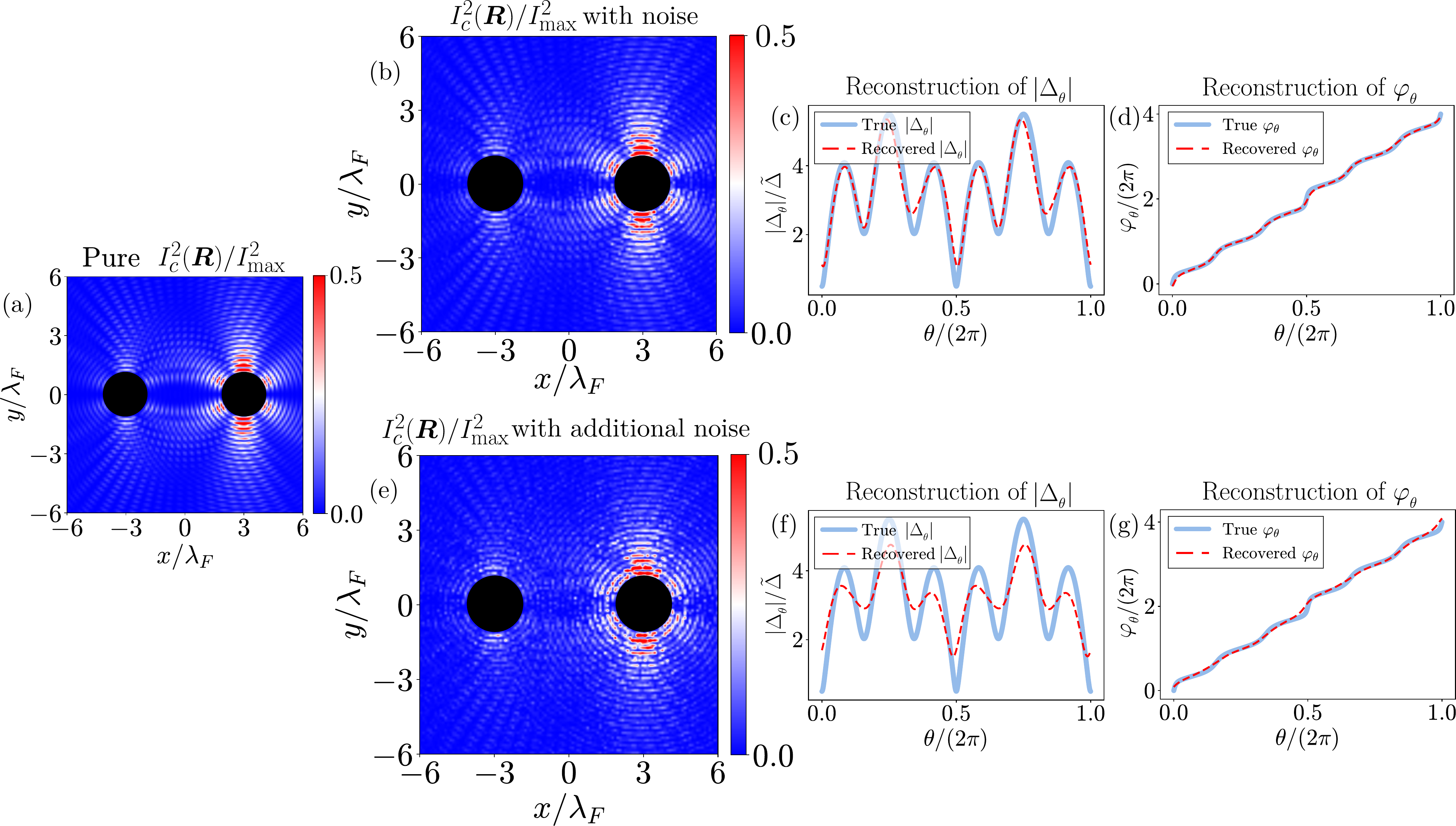}
    \caption{(a) Spatial map of the squared critical current of the fork-tip, where the critical current of the tip 2 is twice stronger than tip 1, i.e., $\eta = (t_2/t_1)^2=2$. Here we have set $\Delta_\theta = \tilde \Delta (-e^{2i\theta } -1.5 e^{-2i\theta} + 3 e^{4 i\theta})$. (b) Here we have introduced some noise $\delta \varphi_i$ to the phase at each point, such that $I_c^2(\vec R) = {I_{0,1}^2(\vec r_1) + I_{0,2}^2(\vec r_2) + 2 I_{0,1}(\vec r_1) I_{0,2}(\vec r_2) \cos((\varphi_1+\delta \varphi_1) - (\varphi_2 + \delta \varphi_2))}$ such that $\delta \varphi_{1}$ and $\delta \varphi_2$ are random Gaussian noise of strength 0.1, computed independently at each spatial point. Panels (c) and (d) show the reconstruction of the angular dependence of magnitude $|\Delta_\theta|$ and phase $\varphi_{\theta}$ of the gap function $\Delta_{\theta}$. Panel (e) shows the spatial pattern of the squared fork-tip critical current with larger Gaussian noise, $\delta \varphi_i$ chosen from random Gaussian noise of strength 0.2. Panels (f) and (g) are analogous to panels (c) and (d), displaying the reconstructed magnitude and phase of the complex gap function. Despite the noise, the fit of the phase winding remains very good because the overall phase winding is fully determined by the leading angular harmonic, and the sub-leading harmonics create small modulations around it.}
    \label{fig:noisy-data}
\end{figure}
When the two tips have unequal tunneling amplitudes $t_1$ and $t_2$, the expression for the total current is modified as,
\begin{equation}
\begin{aligned}
    I_J(\vec{R})&=I_{0,1}(\vec{r}_1)\sin \left(\varphi_1-\varphi_{\mathrm{tip}}\right)+I_{0,2}(\vec{r}_2)\sin \left(\varphi_2-\varphi_{\mathrm{tip}}\right)\\
    &= \sqrt{I_{0,1}^2(\vec r_1) + I_{0,2}^2(\vec r_2) + 2 I_{0,1}(\vec r_1) I_{0,2}(\vec r_2) \cos(\varphi_1 - \varphi_2)} \sin(\alpha - \varphi_{\rm tip})
\end{aligned}
\end{equation}
where the critical current in each tip is given by,
\begin{equation}
    I_{0,i}(\vec r_i)\propto t_i^2 \frac{\sin(2\gamma_{r_i})}{k_F r_i}
T\sum_{i\omega_n}\frac{-|\Delta_{\theta_{\vec r_i}}| }{\sqrt{\omega_n^2+|\Delta_{\theta_{\vec r_i}}|^2}}\frac{|\Delta_{\rm tip}|}{\sqrt{\omega_n^2+|\Delta_{\rm tip}|^2}}.
\end{equation}
and
$\tan(\alpha) =\frac{ I_{0,1}(\vec r_1)\sin(\varphi_1)+ I_{0,2}(\vec r_2)\sin(\varphi_2)}{ I_{0,1}(\vec r_1)\cos(\varphi_1)+I_{0,2}(\vec r_2)\cos(\varphi_2)
}$.

As a result, in the limit of small $\Delta$ such that it can be neglected from the denominator, the fork tip critical current will be proportional to,
\begin{equation}
    I^2_{\text{fork-tip}}(\vec R) \propto |\mathcal F(\vec r_1) + \eta \mathcal F(\vec r_2)|^2 = |(1+\eta)\Delta_0-ug(\vec r_1)\Delta_{\theta_1}-\eta ug(\vec r_2)\Delta_{\theta_2}|^2
\end{equation}
where $\eta = (t_2/t_1)^2$ is the squared ratio of the two tunneling amplitudes.

We set $\eta$ to be an additional unknown parameter in the fitting procedure described in Appendix~\ref{app:curve-fitting}, and the fitted results are shown in Fig.~\ref{fig:noisy-data}.